\documentclass[a4paper,11pt]{article}
\pdfoutput=1 

\usepackage{jheppub} 

\usepackage[T1]{fontenc} 
\usepackage{slashed}

\usepackage{slashbox}

\preprint{arXiv:1604.00847 [hep-th]}

\title{A new algebraic structure in the standard model of particle physics}

\author{Latham Boyle$^1$}
\author{and Shane Farnsworth$^{1,2}$}
\affiliation{$^1$Perimeter Institute for Theoretical Physics, \\ 
Waterloo, Ontario N2L 2Y5, Canada\\
$^2$Max Planck Institute for Gravitational Physics,\\
Am M$\ddot{\text{u}}$hlenberg 1, 14476 Golm, Germany, EU}  

\abstract{We introduce a new formulation of the real-spectral-triple formalism in non-commutative geometry (NCG): we explain its mathematical advantages and its success in capturing the structure of the standard model of particle physics.  The idea, in brief, is to represent $A$ (the algebra of differential forms on some possibly-noncommutative space) on $H$ (the Hilbert space of spinors on that space); and to reinterpret this representation as a simple super-algebra $B=A\oplus H$ with even part $A$ and odd part $H$.  $B$ is the fundamental object in our approach: we show that (nearly) all of the basic axioms and assumptions of the traditional real-spectral-triple formalism of NCG are elegantly recovered from the simple requirement that $B$ should be a differential graded $\ast$-algebra (or ``$\ast$-DGA").  Moreover, this requirement also yields other, new, geometrical constraints.  When we apply our formalism to the NCG traditionally used to describe the standard model of particle physics, we find that these new constraints are physically meaningful and phenomenologically correct.  In particular, these new constraints provide a novel interpretation of electroweak symmetry breaking that is geometric rather than dynamical.  This formalism is more restrictive than effective field theory, and so explains more about the observed structure of the standard model, and offers more guidance about physics beyond the standard model.}

\begin{document}
\maketitle


\section{Introduction}

At low energies, the laws of physics appear to be accurately described by a particular effective field theory (EFT) -- the so-called standard model of particle physics (coupled to Einstein gravity).  However, the EFT framework leaves certain basic questions about the standard model unanswered, and gives us less guidance than we would like about what might come {\it beyond} the standard model.  In the EFT construction of the standard model, one specifies as input the basic symmetries of the theory, as well as the list of fundamental fields and how they transform under those symmetries.  The dynamics are then described by the most general Lagrangian that is built from those fields and is invariant under those symmetries.  However, certain questions remain.  What determines the symmetries -- {\it e.g.}\ why is the standard model gauge group $SU(3)\times SU(2)\times U(1)$?  What determines the basic list of fermions fields, and why do they transform in the particular representations (and with the particular charges) that they do?  Similarly, what determines the basic list of scalar fields, their representations and charges?  It is natural to look for a mathematical framework that is compatible with the EFT description of the standard model, but goes further in addressing some of these questions.  A framework that is more restrictive than EFT -- that only permits a subset of the theories that would seem valid from the EFT standpoint -- could explain more about the standard model, and give more guidance about beyond-the-standard-model physics.    

This paper builds on earlier work on non-commutative geometry (NCG) \cite{ConnesBook, LandiBook, ConnesMarcolliBook, GraciaBondia:2001tr} and its relationship \cite{Connes:1990qp, DuboisViolette:1988ps, DuboisViolette:1989at, DuboisViolette:1988ir, Kastler:1992cu, Kastler:1992ct, Kastler:1995ue, Connes:1995tu, Connes:1996gi, Chamseddine:1991qh, Chamseddine:1996zu, Barrett:2006qq, Connes:2006qv, Chamseddine:2006ep, Chamseddine:2007hz, Chamseddine:2007ia} to the structure of the standard model (for a pedagogical introduction, see \cite{Kastler:2000px, Schucker:2001aa, Chamseddine:2010ps, vandenDungen:2012ky, vanSuijlekomBook}).  The idea, in brief, is that the observed structure of the standard model (coupled to Einstein gravity) may be reinterpreted as arising from the fact that the underlying spacetime is non-commutative ({\it i.e.}\ it is described by a certain kind of NCG, consisting of a 10=4+6 dimensional space, with four continuous/commutative "ordinary" dimensions, and six discrete/non-commutative "extra" dimensions of a certain type).  The claim is that this perspective captures or explains something important about the structure of the standard model that is missed in ordinary EFT.  Much as Euclidean geometry allows one, given the first two angles of a triangle, to infer the third angle, the NCG approach allows one, given certain features of the particle content of a gauge theory, to infer other features of the particle content -- features that would be completely independent inputs from the EFT standpoint.  For example, in the NCG approach, once one chooses: (i) the gauge symmetry, (ii) the basic list of fermion fields in the theory, and (iii) the representations under which they transform, then (iv) the {\it scalar} fields in the theory and the representations under which {\it they} transform are {\it determined} ({\it i.e.}\ they are {\it output}, whereas in EFT they must be specified as additional independent {\it input}).  Fundamentally this is because, in the NCG approach, the Higgs fields have a geometric meaning that puts them on the same footing as the gauge fields: the gauge and Higgs fields are two different pieces of the connection on the non-commutative space.  Moreover, the fermion fields and their representations are much more restricted in NCG than in EFT \cite{Krajewski:1996se, Chamseddine:2007hz, Chamseddine:2007ia}.  Fundamentally this is because in EFT the fermions are governed by the representations of finite-dimensional Lie groups, while in NCG they are governed by the representations of finite-dimensional associative $\ast$-algebras (which are much more restricted: {\it e.g.}\ the Lie group $SU(N)$ has an infinite number of different finite-dimensional irreducible representations, while the associative algebra of $N\times N$ complex matrices $M_{N}(\mathbb{C})$ has only one or two irreducible representations, depending on whether we regard it as an algebra over $\mathbb{C}$ or over $\mathbb{R}$ -- see {\it e.g.}\ \cite{Krajewski:1996se}).   As a simple example, in the NCG approach, the observed fact that all the fermions in the standard model transform in either the trivial or fundamental representation [of $SU(2)$ and $SU(3)$] is an explained {\it output}, whereas in EFT this is an unexplained {\it input}. 

In this paper, we introduce a new formulation of the real-spectral-triple formalism in non-commutative geometry.  We explain its mathematical advantages and its success in capturing the structure of the standard model of particle physics.  Our approach, in brief, is as follows: (i) we start with a $\ast$-algebra $\hat{A}$ (the ``algebra of coordinates"); (ii) we use it to define a related algebra $A$ (the universal $\ast$-algebra of forms over $\hat{A}$); (iii) we take the simplest non-trivial graded representation of $A$ on a Hilbert space $H$ ({\it i.e.}\ we take $H$ to be a graded space with just two non-zero components, $H_{R}$ and $H_{L}$); and (iv) we note that this representation may be reinterpreted as a super-algebra $B=A\oplus H$, with even part $A$ and (square-zero) odd part $H$.  This super-algebra $B$ is the fundamental object in our approach: we note that (nearly) all of the basic axioms and assumptions of the traditional real-spectral-triple" formalism may be elegantly recovered from the simple requirement that $B$ should be a differential graded $\ast$-algebra (or ``$\ast$-DGA").  In addition, this requirement also yields other, new, geometrical constraints.  When we apply our formalism to the spectral triple traditionally used to describe the geometry of the standard model of particle physics, we find that these new constraints are physically meaningful and phenomenologically correct (although a number of puzzles remain -- see Section \ref{Discussion}).  Notably, these new constraints give a new interpretation/explanation of electroweak symmetry breaking that is geometric/algebraic, rather than dynamical.

In a sense, our new proposal only differs from our earlier one \cite{Boyle:2014wba, Farnsworth:2014vva} by a small change -- namely, the fact that the representation of $A$ on $H$ is now appropriately graded --  but this small change leads to many improvements.
Although (as reviewed in Section 6 below) our earlier proposal already had some of the nice features of the new one, it also had two key drawbacks (mentioned in \cite{Boyle:2014wba}, but emphasized and clarified in \cite{Brouder:2015qoa}).  The first drawback was that, although $A$ was a differential graded algebra, its extension to $B$ was not (due to the presence of junk forms \cite{ConnesLottJunk, LandiBook, ConnesMarcolliBook}).  The second drawback was that, although our new "second-order condition" seemed to mesh very nicely with the six-dimensional discrete/non-commutative part of the standard model geometry, it was incompatible with the ordinary four-dimensional continuous/commutative part.  In their paper \cite{Brouder:2015qoa}, Brouder {\it et al} had an important insight -- they pointed out that these two problems could be simultaneously resolved by taking an appropriately {\it graded} representation of $A$ on $H$.  Although the formulation we suggest here is very different than the one proposed by Brouder {\it et al}, their basic insight -- that the representation of $A$ on $H$ should be appropriately graded -- is still the key.  In this paper,
 we point out a different (more minimal) graded representation; and we find that many other pieces then fall neatly into place.

The outline of this paper is as follows.  In Section 2 we introduce the idea of a differential-graded $\ast$-algebra (or "$\ast$-DGA").  Although $\ast$-algebras and (non-commutative) DGAs have been widely studied on their own, the combined object (a non-commutative $\ast$-DGA) seems to have been studied less, and has some novel features that will play a key role in our analysis.  We particularly direct the reader's attention to remarks (ii)$'$, (v)$'$ and (v)$''$, which do not seem widely known, and will be crucial in what follows.  The goal of Section 3 is: first, to introduce Eilenberg's idea that the represention of an algebra $A$ may be regarded as a new "Eilenberg algebra" $B=A\oplus H$ (a ``square-zero extension" of $A$, and a particularly simple type of super-algebra); second, to explain how this idea naturally generalizes to representing $\ast$-algebras, DGAs, and $\ast$-DGAs; and third, to define the tensor product of two Eilenberg $\ast$-DGAs.   In Section 4, we introduce a simple type of Eilenberg $\ast$-DGA $B=A\oplus H$ and explain its relevance to NCG.  First (in Subsection \ref{defining_A}), given a $\ast$-algebra $\hat{A}$ (the "algebra of coordinates"), we define a corresponding algebra $A$ (the universal $\ast$-algebra of differential forms over $\hat{A}$). Second (in Subsection \ref{defining_B}), we take the simplest non-trivial graded representation of $A$ on a Hilbert space $H$ ({\it i.e.}\ we take $H$ to be a graded space with just two non-zero components $H=H_{L}\oplus H_{R}$).  Third (in Subsection \ref{NCG_comparison}), we show that the simple and natural requirement that $B$ is an associative $\ast$-DGA elegantly unifies (nearly) all of the axioms and assumptions underlying the traditional real-spectral-triple formalism of NCG; and, in addition, we show that this same requirement also yields other, new, geometric constraints (which do not appear in the traditional formulation).  In Section 5, we apply our formalism to the particular geometric data traditionally used to describe the standard model in NCG; and we show that our new geometric constraints correspond to physically meaningful and phenomenologically correct conditions (which thereby represent an improvement over the traditional spectral triple formulation of the standard model, although there are remaining puzzles which indicate that the story still has missing pieces).  In Section 6, we review some phenomenological consequences of our previous formulation that carry over to the new formulation, and we mention some puzzles and interesting directions for future work.    

\section{Differential graded $\ast$-algebras ("$\ast$-DGAs")}
\label{dga}

We begin, in this section, with six definitions: (i) an algebra; (ii) a $\ast$-algebra; (iii) a graded algebra; (iv) a differential graded algebra (or "DGA"); (v) a differential graded $\ast$-algebra (or "$\ast$-DGA"); and (vi) the tensor product of two $\ast$-DGAs.  

Although $\ast$-algebras and (non-commutative) DGAs have been widely studied on their own, the combined object (a non-commutative $\ast$-DGA) seems to have been studied less, and has some novel features that will play a key role in our analysis.  We particularly direct the reader's attention to remarks (ii)$'$, (v)$'$ and (v)$''$, which do not seem widely known, and will be crucial in what follows.

(i) An algebra $A$ (over a field $\mathbb{F}$) is a vector space (over $\mathbb{F}$) equipped with an $\mathbb{F}$-bilinear product $aa'\in A$ (for $a,a'\in A$).  $A$ is called "commutative" if the "commutator" $[a,a']\equiv aa'-a'a$ vanishes for all $a,a'\in A$; and "associative" if the  "associator" $[a,a',a'']\equiv (aa')a''-a(a'a'')$ vanishes for all $a,a',a''\in A$.  (Note that we do not assume either commutativity or associativity of $A$ in this section.)

(ii) A $\ast$-algebra is an algebra $A$ (over $\mathbb{F}$) that is equipped with an additional structure: a $\ast$ operation.  By a $\ast$ operation, we mean an anti-automorphism from $A$ to $A$ -- {\it i.e.}\ an invertible $\mathbb{F}$-anti-linear map from $A$ to $A$ that is an anti-homomorphism:
\begin{equation}
  \label{star}
  (aa')^{\ast}=a'{}^{\ast}a^{\ast}.
\end{equation}
and also has the property that $(a^{\ast})^{\ast}$ is proportional to $a$:
\begin{equation}
  \label{epsilon}
  (a^{\ast})^{\ast}=\epsilon a.
\end{equation}
For example: the complex numbers $\mathbb{C}$ are a commutative $\ast$-algebra (over $\mathbb{R}$), where the $\ast$ operation is complex conjugation $z\to\bar{z}$; and $n\times n$ complex matrices $M_{n}(\mathbb{C})$ are a non-commutative $\ast$-algebra (over $\mathbb{C}$), where the $\ast$ operation is the conjugate transpose ($\dagger$) operation $m\to m^{\dagger}$.

(ii)$'$ Ordinarily, we can then use the argument $\epsilon aa'=((aa')^{\ast})^{\ast}=(a'{}^{\ast}a^{\ast})^{\ast}=(a^{\ast})^{\ast}(a'{}^{\ast})^{\ast}
=\epsilon^{2} aa'$ to determine that $\epsilon=+1$, so that $\ast$ becomes an involution, $(a^{\ast})^{\ast}=a$; but, as we will see in the 
Subsection \ref{representing_star}, this argument fails in the nilpotent sector of an Eilenberg algebra, and the more general possibility $\epsilon=\pm1$ is allowed.\footnote{We have chosen the name $\epsilon$ for this $\pm$ sign because, as we will see in Section \ref{NCG}, it is linked to the $\pm$ sign $\epsilon$ that appears in the standard definition of a real spectral triple in NCG: $J^{2}=\epsilon$ (see \cite{vandenDungen:2012ky, vanSuijlekomBook}).}  In the general case where $(a^{\ast})^{\ast}\neq a$, we should be careful to distinguish between the operation "$\ast$" and its inverse "$\bar{\ast}$".

(iii) A graded algebra $A$ is an algebra that decomposes into subspaces, $A=\oplus_{m}A_{m}$, where the product respects the decomposition: $a_{m}\in A_{m}$, $a_{n}\in A_{n}$ $\Rightarrow$ $a_{m}a_{n}\in A_{m+n}$.    

(iv) A differential graded algebra (or "DGA") is a graded algebra $A$ that is also equipped with a left-differential $d_{L}^{}$: a linear map from $A_{m}$ to $A_{m+1}$ that is nilpotent 
\begin{subequations}
  \begin{equation}
    \label{dL_nilpotent}
    d_{L}^{2}=0
  \end{equation}
  and satisfies the left-Leibniz rule (see Appendix \ref{graded_Leibniz_rule})
  \begin{equation}
    \label{graded_Leibniz}
    d_{L}^{}(a_{m}^{}a_{n}^{})=d_{L}^{}(a_{m}^{})a_{n}^{}+(-1)^{m}a_{m}^{}d_{L}^{}(a_{n}^{})\qquad(a_{m}^{}\in A_{m}^{}, a_{n}^{}\in A_{n}^{}).
  \end{equation}
\end{subequations}
An example is the exterior algebra of differential forms: it is graded, since an $m$-form wedged with an $n$-form is an $(m+n)$-form, and it is equipped with a differential: the usual exterior derivative $d$ on differential forms.

(v) A differential graded $\ast$-algebra (or "$\ast$-DGA") is a DGA that is also a $\ast$-algebra (with the properties listed above).  In particular, Eq.~(\ref{star}) becomes (see Appendix \ref{graded_Leibniz_rule})
\begin{equation}
  \label{graded_star}
  (a_{m}^{}a_{n}^{})^{\ast}=a_{n}^{\ast}a_{m}^{\ast}.
\end{equation} 

(v)$'$ The $\ast$ operation maps $m$-forms to $f(m)$-forms; here Eq.~(\ref{graded_star}) implies $f(m+n)=f(m)+f(n)$ (which implies $f(m)=\epsilon'' m$, for some constant $\epsilon''$), while Eq.~(\ref{epsilon}) implies $f(f(m))=m$ (which implies $\epsilon''=\pm1$).   This means that $\ast$-DGAs naturally come in two distinct flavors: the $\epsilon''=+1$ flavor, where the $\ast$ operation maps $m$-forms to $m$-forms, and the $\epsilon''=-1$ flavor, where the $\ast$ operation maps $m$-forms to $(-m)$-forms.\footnote{We have chosen the name $\epsilon''$ for this $\pm$ sign because, as we will see in Section \ref{NCG}, for the models we consider it is linked to the sign $\epsilon''$ that appears in the standard definition of a real even spectral triple in NCG: $J\gamma=\epsilon''\gamma J$ (see \cite{vandenDungen:2012ky, vanSuijlekomBook}). Care should be taken however, as this correspondence is not true in general, and only holds for NCG models in which the $\mathbb{Z}_2$ grading of an even spectral triple is identified with the differential grading of its corresponding Eilenberg $*$-DGA (as explained below in Sections~\ref{Fused} and~\ref{NCG}).}  This observation, although very basic, will be crucial in what follows (and may even be novel).   

(v)$''$ A $\ast$-DGA comes equipped with a left-differential $d_{L}$ and a $\ast$-operation; and from these, it is natural to also construct the right-differential $d_{R}$:
\begin{equation}
  d_{R}^{}\equiv \ast\circ d_{L}^{}\circ\bar{\ast}\label{dga_dR}
\end{equation}
({\it i.e.} "$\ast$" composed with $d_{L}$ composed with "inverse $\ast$").  Note that, since $d_{L}$ is nilpotent (\ref{dL_nilpotent}) and satisfies the left-Leibniz rule (\ref{graded_Leibniz}), it follows that $d_{R}$ is also nilpotent:
\begin{subequations}
  \begin{equation}
    d_{R}^{2}=0
  \end{equation}
  and satisfies the right-Leibniz rule (see Appendix \ref{graded_Leibniz_rule})
  \begin{equation}
    d_{R}(a_{m}^{}a_{n}^{})=a_{m}^{}d_{R}(a_{n}^{})+(-1)^{n}d_{R}(a_{m}^{}) a_{n}^{}.
  \end{equation}
\end{subequations}
Note that $d_{R}$ maps $m$-forms to $(m+\epsilon'')$ forms: in other words, when $\epsilon''=+1$, it is a right-differential in the usual sense, but when $\epsilon''=-1$, it is a right-differential with respect to the inverted grading.  Also note the following identities which follow from the definition of $d_{R}$
\begin{subequations}
  \begin{eqnarray}
    d_{L\;\!}^{}(a_{m}^{\ast})&=&d_{R}^{}(a_{m}^{})^{\ast}, \\
    d_{R}^{}(a_{m}^{\ast})&=&d_{L\;\!}^{}(a_{m}^{})^{\ast},
  \end{eqnarray}
\end{subequations}
and which say that a $\ast$ operation which appears inside the argument of a differential may be "pulled outside" at the cost of swapping 
$d_{L}\leftrightarrow d_{R}$.

(vi) Given two $\ast$-DGAs $(A',d')$ and $(A'',d'')$, their tensor product $(A,d)$ is defined as follows: the vector space $A$ is the tensor product of the vector spaces $A'$ and $A''$ ($A= A'\otimes A''$), the product on $A$ is given by (see Appendix \ref{graded_product_rule}): 
\begin{equation}
  \label{tensor_product}
  (a_{m}'\otimes a_{n}'')(a_{p}'\otimes a_{q}'')=(-1)^{np}a_{m}'a_{p}'\otimes a_{n}''a_{q}'',
\end{equation}
the $\ast$ operation on $A$ is given by (see Appendix \ref{graded_product_rule}):
\begin{equation}
  \label{tensor_star}
  (a_{m}'\otimes a_{n}'')^{\ast}=(-1)^{mn}a_{m}'{}^{\ast}\otimes a_{n}''{}^{\ast},
\end{equation}
and the differential $d$ is given by:
\begin{equation}
  \label{tensor_d}
  d=d'\otimes 1''+1'\otimes d''
\end{equation}
where $1'$ and $1''$ denote the identity operators on $A'$ and $A''$, respectively.

\section{Representing $\ast$-DGAs}
\label{Fused}

The goal of this section is to introduce Eilenberg's idea that the represention of an algebra $A$ may be regarded as a new "Eilenberg algebra" (a particular type of super-algebra with $A$ as its even part); to explain how this idea naturally generalizes to representing $\ast$-algebras, DGAs, and $\ast$-DGAs;  and to define the tensor product of two Eilenberg $\ast$-DGAs.   

\subsection{Representing algebras (with Eilenberg algebras)}
\label{representing_algebras}

In this subsection, we introduce Eilenberg's perspective on representing an algebra $A$ (via an associated super-algebra $B$).

Let $A$ be an algebra (over $\mathbb{F}$), and let $H$ be a vector space (over $\mathbb{F}$); following Eilenberg and Schafer \cite{Eilenberg, Schafer}, we define a bi-representation $R$ of $A$ on $H$ (or, equivalently, a bi-module $H$ over $A$) as a pair of $\mathbb{F}$-bilinear products $ah\in H$ and $ha\in H$ ($a\in A$, $h\in H$).

Now notice that this definition of a bi-representation of $A$ on $H$ (or, equivalently, of a bi-module $H$ over $A$) is equivalent to the definition of a new algebra
\begin{equation}
  \label{B}
  B=A\oplus H,
\end{equation}
(over $\mathbb{F}$) with the product between elements of $B$ ($b=a+h$ and $b'=a'+h'$) given by
\begin{equation}
  \label{B_product}
  bb'=aa'+ah'+ha'
\end{equation}
where $aa'\in A$ is the product inherited from $A$, while $ah'\in H$ and $ha'\in H$ are the products inherited from $R$, and $hh'=0$.  
We will call such an algebra an "Eilenberg algebra."\footnote{This name was introduced in Ref.~\cite{Brouder:2015qoa}. It is worth noting that the definition of an Eilenberg algebra originally introduced in \cite{Eilenberg} and reviewed in \cite{Brouder:2015qoa} is slightly more elaborate and general than the simpler definition presented in \cite{Schafer, Boyle:2014wba, Farnsworth:2014vva} and adopted in the present paper.  We mention this in case the greater generality afforded by Eilenberg's original formulation turns out to be important for future developments of the formalism presented here.  This is also closely related to the idea of a `square-zero extension'~\cite{Weibel1994book}.}

Also notice that the algebra $B$ defined this way is automatically a superalgebra -- {\it i.e.}\ a $\mathbb{Z}_{2}$-graded algebra, with ``even" and ``odd" subspaces $A$ and $H$, respectively.  

We stress that, so far, we have not assumed anything about the associativity of $A$ or $B$.  On the one hand, {\it if} we now assume that $B$ is associative, then we precisely recover the traditional associative definition of an ordinary (left-, right-, or bi-)representation of the algebra $A$ on $H$, in the sense described in the following paragraph.  On the other hand, we need not necessarily assume that $B$ is associative: for example, if $A$ is a Jordan algebra (an important type of non-associative algebra), then it is natural to define its representation on $H$ by taking $B$ to also be a Jordan algebra \cite{Eilenberg, Schafer, Jacobson}.  In fact, this is what originally led us to adopt Eilenberg's perspective in \cite{Farnsworth:2013nza, Boyle:2014wba}: it is a way of defining the representation of $A$ on $H$ that naturally generalizes from non-commutative geometry (where the algebra of coordinates, may be non-commutative) to non-associative geometry (where the algebra of coordinates may also be non-associative).

Let us now explain our assertion (from the previous paragraph) that if we assume $B$ is associative, then we precisely recover the usual associative definition of an ordinary representation of the algebra $A$ on $H$.  If $B$ is associative, all the associators $[b,b',b'']$ must vanish.  This implies four non-trivial constraints:
\begin{subequations}
  \begin{eqnarray}
    \label{aaa}
    [a,a',a'']&=&0, \\
    \label{aah}
    [a,a',h'']&=&0, \\
    \label{haa}
    [h,a',a'']&=&0, \\
    \label{aha}
    [a,h',a''] &=&0, 
  \end{eqnarray}
\end{subequations}
while the remaining associators (in which two or three arguments are from $H$) vanish trivially because $hh'=0$.  Note that (\ref{aaa}) is simply the requirement that $A$ itself is associative; (\ref{aah}) says that $ah$ is a traditional associative left-representation of $A$ on $H$; (\ref{haa}) says that $ha$ is a traditional associative right-representation of $A$ on $H$; and (\ref{aha}) says that the left- and right-representations commute with each other.  In other words, we recover the traditional associative definition of a left-right bi-representation of $A$ on $H$ (or, equivalently, the traditional associative definition of a left-right bi-module $H$ over $A$); and the special cases of a left-representation (left-module) or right-representation (right-module) are recovered, respectively, when either the right action $ha$ or the left action $ah$ vanishes identically.

\subsection{Representing $\ast$-algebras (with Eilenberg $\ast$-algebras)}
\label{representing_star}

In this subsection, we explain how to extend Eilenberg's idea from algebras to $\ast$-algebras: one simply requires that $B$ itself is a $\ast$-algebra.  We will see how many of the traditional axioms/assumptions of NCG follow from this requirement.   

To extend Eilenberg's construction from algebras to $\ast$-algebras, we must promote $B$ from an algebra to a $\ast$-algebra.  The $\ast$ operation on $B$ should have all the properties explained in Section \ref{dga}, plus one more: compatibility with the $\ast$ operation on the sub-algebra $A\subset B$. This together with~\eqref{epsilon} implies compatibility with the intrinsic $\mathbb{Z}_{2}$ grading on $B$ ({\it i.e.}  $b^{\ast}\in A$ when $b\in A$, and $b^{\ast}\in H$ when $b\in H$), and fixes the $\ast$ operation to be
\begin{equation}
  \label{B_star}
  b^{\ast}=a^{\ast}+ J h
\end{equation}
for $a\in A$, $h\in H$, and where $a^{\ast}$ is the $\ast$-operation on $A$, while $J$ is an invertible anti-linear operator on $H$.  We will call such an algebra an "Eilenberg $\ast$-algebra."

In Section \ref{dga}, we took the $\ast$-algebra $A$ to satisfy $(a^{\ast})^{\ast}=\epsilon a$ (and then derived $\epsilon=1$); but for an Eilenberg super-algebra we should allow a different constant in the even and odd sectors: $(a^{\ast})^{\ast}=\epsilon_{0}^{}a$ and $(h^{\ast})^{\ast}=\epsilon_{1}^{}h$.  Then, as in Section \ref{dga}, the argument $\epsilon_{0}^{}aa'=((aa')^{\ast})^{\ast}=(a'{}^{\ast}a^{\ast})^{\ast}=(a^{\ast})^{\ast}(a'{}^{\ast})^{\ast}=\epsilon_{0}^{2}aa'$ yields $\epsilon_{0}^{}=1$; but $\epsilon_{1}^{}ah=((ah)^{\ast})^{\ast}=(h^{\ast}a^{\ast})^{\ast}=(a^{\ast})^{\ast}(h^{\ast})^{\ast}=\epsilon_{1}ah$ doesn't yield any constraint on $\epsilon_{1}^{}$; and $\epsilon_{1}^{}hh'=((hh')^{\ast})^{\ast}=(h'{}^{\ast}h^{\ast})^{\ast}=(h^{\ast})^{\ast}(h'{}^{\ast})^{\ast}=\epsilon_{1}^{2}hh'$ doesn't either (because $hh'=0$).  If the $\ast$ operation still has {\it some} finite period ({\it i.e.} $\ast^{n}=1$ for some finite $n$): it follows that $\epsilon_{1}$ is a root of unity; but then, the argument $\epsilon_{1}h^{\ast}=(h^{\ast})^{\ast\ast}=(h^{\ast\ast})^{\ast}=(\epsilon_{1}h)^{\ast}=\bar{\epsilon}_{1}h^{\ast}$
implies that $\epsilon_{1}$ is also real, and hence $\pm1$.  From now on, to match standard NCG notation, let us drop the subscript "$1$" and simply refer to $\epsilon_{1}$ as "$\epsilon$."  We thus recover the standard NCG axiom $J^{2}=\epsilon$, where $\epsilon=\pm1$.

The fact that $B$ is a $\ast$-algebra thus implies and unifies four traditionally-assumed facts about NCG, including: (i) that $A$ is a $\ast$-algebra; (ii) that $H$ is equipped with an invertible anti-linear operator $J$; and (iii) that $J^{2}=\epsilon$ where $\epsilon=\pm1$.  In addition, (iv) the anti-homomorphism property $(bb')^{\ast}=b'{}^{\ast}b^{\ast}$ implies $(ah)^{\ast}=h^{\ast}a^{\ast}$ and $(ha)^{\ast}=a^{\ast}h^{\ast}$, which then implies that $A$ is not just left-represented or right-represented on $H$, but left-right bi-represented on $H$, with the left and right representations related by
\begin{subequations}
  \label{RL_relations}
  \begin{eqnarray}
    R_{a}&=&J L_{a^{\ast}}J^{-1}, \\
    L_{a}&=&J R_{a^{\ast}}J^{-1}.
  \end{eqnarray}
\end{subequations}

Finally, in the NCG context, $H$ will be a Hilbert space, so compatibility with the inner product on $H$ will also require $J$ to be anti-unitary
\begin{equation}
  J^{\dagger}=J^{-1}.
\end{equation}

\subsection{Representing DGAs (with Eilenberg DGAs)}
\label{representing_dga}

In this subsection, we explain how to extend Eilenberg's idea from algebras to DGAs: in parallel with the previous subsections, one now requires that $B$ itself is a DGA.  

Let us proceed in two steps: (i) first, we define graded bi-representations (or, equivalently, graded bi-modules), and (ii) second we define differential graded bi-representations (or, equivalently, differential graded bi-modules).

(i) Suppose $A$ is a graded algebra (over $\mathbb{F}$) with grading $A=\oplus_{m}A_{m}$, and $H$ is a vector space (over $\mathbb{F}$) with grading $H=\oplus_{m}H_{m}$.  As the natural extension of Eilenberg's preceding definition of a bi-representation (or bi-module), let us define a {\it graded} bi-representation $R$ of $A$ on $H$ (or, equivalently, a {\it graded} bi-module $H$ over $A$) as a pair of $\mathbb{F}$-bilinear products $ah\in H$ and $ha\in H$ that respect the grading:
\begin{equation}
  a_{m}h_{n}\in H_{m+n}\quad{\rm and}\quad h_{n}a_{m}\in H_{m+n}\qquad(a_{m}\in A_{m},\;h_{n}\in H_{n}). 
\end{equation}

(ii) Next let $A$ be a DGA, with grading $A=\oplus_{m}A_{m}$ and differential $d$; and let $H$ be a vector space with grading $H=\oplus_{m}H_{m}$.  Then we will say that a graded bi-representation $R$ of $A$ on $H$ is a {\it differential} graded bi-representation of $A$ on $H$ (or, equivalently, a {\it differential} graded bi-module $H$ over $A$) if $H$ is also equipped with its own differential $d$: {\it i.e.}\ a linear operator from $H_{m}$ to $H_{m+1}$ that is nilpotent 
\begin{equation}
  \label{B_nilpotent}
  d^{2}(h_{n})=0
\end{equation}
and satisfies the graded Leibniz conditions
\begin{subequations}
  \label{graded_Leibniz_Eilenberg}
  \begin{eqnarray}
    d(a_{m}h_{n})&=&d(a_{m})h_{n}+(-1)^{m}a_{m}d(h_{n}), \\
    d(h_{n}a_{m})&=&d(h_{n})a_{m}+(-1)^{\;\!n\;\!}h_{n}d(a_{m}),
  \end{eqnarray}
\end{subequations}
for $a_{m}\in A_{m}$ and $h_{n}\in H_{n}$.

With these definitions, we now notice that the definition of a graded bi-representation of the graded algebra $A$ on $H$ (or, equivalently, the definition of a graded bi-module $H$ over $A$) is simply equivalent to the definition of the algebra $B$ given above, together with the condition that $B$ is graded as follows:
\begin{equation}
  \label{B_grading}
  B=\bigoplus_{m}(A_{m}\oplus H_{m})
\end{equation}
[where in the applications to NCG in Section~\ref{NCG}, $A_m$ will only be non-zero for integer values of $m$ or some subset thereof, while $H_m$ will only be non-zero for $m = \{0,1\}$ (when $\epsilon'' = 1$) or for $m=\pm \tfrac{1}{2}$ (when $\epsilon'' = -1$)]. Similarly, the definition of a differential graded bi-representation of the DGA $A$ on $H$ (or, equivalently, the definition of a differential graded bi-module $H$ over $A$) is simply equivalent to the requirement that the algebra $B$ is itself a DGA with respect to the grading (\ref{B_grading}).  We will call such an algebra $B$ an "Eilenberg DGA."

We also notice that the algebra $B$ is now graded in two different ways or, more precisely, it is graded over the ring $\mathbb{Z}\times\mathbb{Z}_{2}$.  In other words, it has its new grading $\bigoplus_{m}(A_{m}\oplus H_{m})$ over $\mathbb{Z}$; but, in addition, it still has another independent $\mathbb{Z}_{2}$ grading that splits it into an even part $A$ and an odd part $H$, thereby making it a super-algebra (or, in this case, a super-DGA).

We again stress that, so far in this subsection, we have not assumed anything about the associativity of $A$ or $B$.  {\it If} we now assume that $B$ is associative, then we recover a case of the traditional associative definition of a (differential) graded representation of the (differential) graded algebra $A$ on $H$.  However, as discussed above, we need not necessarily assume that $B$ is associative; and for some purposes, the fact that this approach to representing a (differential) graded algebra naturally generalizes to the non-associative case may be crucial.\footnote{{\it e.g.}\ in the generalization from non-commutative to non-associative geometry; and perhaps also for describing beyond-the-standard-model physics.}

\subsection{Representing $\ast$-DGAs (with Eilenberg $\ast$-DGAs)}
\label{representing_star_dga}

Finally, in this subsection, we extend Eilenberg's idea from algebras to $\ast$-DGAs (by requiring $B$ itself is a $\ast$-DGA).  

Having laid all the groundwork in the previous three subsections, there is little to be added here.  A differential graded $\ast$-representation of $A$ on $H$ (or, equivalently, a differential graded $\ast$-module $H$ over $A$) is simultaneously a $\ast$-representation ($\ast$-module) in the sense of Subsection \ref{representing_star}, and a differential graded representation (differential graded module) in the sense of Subsection \ref{representing_dga}.  As before, these conditions may be succinctly summarized by saying that the algebra $B$ defined above is a $\ast$-DGA (and, moreover, a super-$\ast$-algebra, because of its intrinsic $\mathbb{Z}_{2}$ grading).  We will call such an algebra an "Eilenberg $\ast$-DGA."

\subsection{Tensoring two Eilenberg $\ast$-DGAs}
\label{tensoring_Eilenberg}

We have seen how a $\ast$-DGA $A$ may be represented by an Eilenberg $\ast$-DGA $B=A\oplus H$; and we would now like to define the tensor product of two such Eilenberg $\ast$-DGAs.  At the end of Section \ref{dga}, we explained how to take the tensor product of two generic $\ast$-DGAs $A'$ and $A''$ to obtain a new $\ast$-DGA $A$.  If we directly apply this construction to two Eilenberg DGAs $B'=A'\oplus H'$ and $B''=A''\oplus H''$, we obtain a new $\ast$-DGA $\hat{B}$; but it is not an {\it Eilenberg} $\ast$-DGA, since its four components $\hat{B}=(A'\otimes A'')\oplus(A'\otimes H'')\oplus(H'\otimes A'')\oplus(H'\otimes H'')$ do not decompose into two pieces $A\oplus H$ with the necessary properties: $A^{2}\in A$, $AH\in H$, $HA\in H$ and $H^{2}=0$.  The remedy is to simply throw away the "odd" parts $A'\otimes H''$ and $H'\otimes A''$ in $\hat{B}$: the remaining even sub-algebra $B=(A'\otimes A'')\oplus(H'\otimes H'')$ (which is indeed an Eilenberg $\ast$-DGA) is the correct definition for the product of the two Eilenberg $\ast$-DGAs $B'$ and $B''$.  Note a somewhat confusing point: the "even" sub-algebra $B=(A'\otimes A'')\oplus(H'\otimes H'')$ is, itself, $\mathbb{Z}_{2}$-graded (like all Eilenberg algebras), and hence breaks into an even part $A=A'\otimes A''$ and odd part $H=H'\otimes H''$, with respect to this grading.

\section{Application to NCG}
\label{NCG}

In the previous section we explained, in general terms, how to extend Eilenberg's approach: from representing algebras to representing $\ast$-DGAs.  In this section, our goal is to describe a simple type of Eilenberg $\ast$-DGA $B=A\oplus H$: first, given a $\ast$-algebra $\hat{A}$ (the "algebra of coordinates"), we define $A$ as the universal $\ast$-algebra of differential forms over $\hat{A}$; and then, to complete the definition of $B$, we take $H$ to be the simplest possible non-trivially graded space -- a space with just two components $H=H_{L}\oplus H_{R}$ -- and follow this idea where it leads.  We find that this construction does a remarkably good job of unifying and explaining many aspects of the traditional NCG formalism\footnote{Note that in this paper we focus on NCG spaces of even KO dimension and Euclidean signature, leaving spaces of odd dimension and Lorentzian signature to future work.}, on the one hand, and resolving key problems/puzzles in the traditional NCG construction of the standard model of particle physics, on the other.  

So far in this paper, we have not assumed associativity, since we want to emphasize that one of the features of our formalism is that it retains the advantage of Eilenberg's perspective -- {\it i.e.}\ it naturally lends itself to generalization, from the associative to the non-associative case.  This is surely an interesting direction for future work.  However, for the rest of this paper, we will restrict to the associative case -- {\it i.e.} the case where $B$ is associative (and hence $A$ and $\hat{A}$ are also associative -- since our main goal in the remainder of the paper will be to show how we can thereby neatly unify and illuminate the traditional axioms of traditional (associative) NCG and also fix several problems in the traditional NCG construction of the standard model.  

\subsection{The $\ast$-DGA $A$}
\label{defining_A}

As in the traditional real-spectral-triple formalism of NCG, we start by choosing a (possibly non-commutative) $\ast$-algebra $\hat{A}$ (over $\mathbb{F}$): roughly, this may be thought of as the algebra of coordinates.  We can then define $A$, the universal $\ast$-algebra of forms over $\hat{A}$, as follows:  

For every element $\hat{a}\in\hat{A}$, let us introduce a corresponding formal symbol $d(\hat{a})$ which has the following familiar linearity and Leibniz properties: $d(\lambda\hat{a})=\lambda d(\hat{a})$, $d(\hat{a}+\hat{a}')=d(a)+d(\hat{a}')$, $d(\hat{a}\hat{a}')=d(\hat{a})\hat{a}'+\hat{a}d(\hat{a}')$ ($\lambda\in\mathbb{F}$, $\hat{a},\hat{a}'\in\hat{A}$).  We can regard $\hat{a}$ as a zero-form, $d(\hat{a})$ as a one-form, and $d(\hat{a})^{\ast}$ as an $\epsilon''$-form ({\it i.e.}\ as a one-form or a minus-one-form, depending on the sign of $\epsilon''$ -- see Section \ref{dga}).  Next consider an arbitrary term constructed by taking a product of $\hat{a}$'s, $d(\hat{a})$'s and $d(\hat{a})^{\ast}$'s: to take a rather complicated example, consider $\hat{a}^{(1)}d(\hat{a}^{(2)})^{\ast}d(\hat{a}^{(3)})\hat{a}^{(4)}d(\hat{a}^{(5)})\hat{a}^{(6)}$, where $\hat{a}^{(1)},\ldots,\hat{a}^{(6)}\in\hat{A}$: note that this example is a 3-form if $\epsilon''=+1$, or a 1-form if $\epsilon''=-1$.  We can take the product of two such terms by simply juxtaposing them, and using the product inherited from $\hat{A}$: {\it e.g.}, the product of the 1-form $a_{1}^{}=d(\hat{a}^{(1)})\hat{a}^{(2)}$ and the one-form $a_{1}'=\hat{a}^{(3)}d(\hat{a}^{(4)})$ is the 2-form $a_{2}=a_{1}^{}a_{1}'=d(\hat{a}^{(1)})\hat{a}^{(5)}d(\hat{a}^{(4)})$, where we have used the product $\hat{a}^{(2)}\hat{a}^{(3)}=\hat{a}^{(5)}$ in $\hat{A}$.  The algebra $A$ is obtained by considering all such terms (and all linear combinations of such terms over $\mathbb{F}$), with the product just defined.  Note that this algebra is automatically graded, $A=\oplus_{m}A_{m}$, where $A_{m}$ is the space of $m$-forms.  The $\ast$ operation on $\hat{A}$ extends to a $\ast$ operation on $A$ in the natural way (by recursive application of the rules described in Section \ref{dga}): so, for example, $(\lambda\,d(\hat{a}^{(1)})\,\hat{a}^{(2)}{}^{\ast}\,d(\hat{a}^{(3)})^{\ast}\,\hat{a}^{(4)})^{\ast}=\bar{\lambda}\,\hat{a}^{(4)}{}^{\ast}\,d(\hat{a}^{(3)})\,\hat{a}^{(2)}\,d(\hat{a}^{(1)})^{\ast}$.  We can also extend $d(\ldots)$ to a differential on $A$ in the natural way by recursive application of the graded Leibniz rule $d(a_{m}a_{n})=d(a_{m})a_{n}+(-1)^{m}a_{m}d(a_{n})$ ($a_{m}\in A_{m}$, $a_{n}\in A_{n}$), along with the definitions $d(d(\hat{a}))=d(d(\hat{a})^{\ast})=0$ ($\hat{a}\in \hat{A}$): so, for example we can write $d(\hat{a}^{(1)}d(\hat{a}^{(2)})\hat{a}^{(3)}d(\hat{a}^{(4)})^{\ast})=d(\hat{a}^{(1)})d(\hat{a}^{(2)})\hat{a}^{(3)}d(\hat{a}^{(4)})^{\ast}-\hat{a}^{(1)}d(\hat{a}^{(2)})d(\hat{a}^{(3)})d(\hat{a}^{(4)})^{\ast}$.  In short: $A$, the universal $\ast$-algebra of differential forms over $\hat{A}$, is the algebra generated by $\hat{A}$, $d(\hat{A}\;\!)$ and $d(\hat{A}\;\!)^{\ast}$, modulo the various relations described above.  

Note that this "universal $\ast$-algebra of differential forms over $\hat{A}$" is similar to the more familiar "universal algebra of differential forms over $\hat{A}$" (presented {\it e.g.}\ in \cite{LandiBook}).  However, as indicated by the difference in name, there is an important difference between these two algebras, coming from the way the $\ast$ operation is handled: note, in particular, that we do not add the extra relation $d(\hat{a})^{\ast}=-d(\hat{a}^{\ast})$ (or $d(\hat{a})^{\ast}=+d(\hat{a}^{\ast}])$), but instead  only impose  that  $d(\hat{a})^{\ast}$ is an $\epsilon''$-form, with the weaker condition $d(d(\hat{a})^{\ast})=0$.  This difference will be key in what follows.  

\subsection{The $\ast$-DGA $B=A\oplus H$}
\label{defining_B}

So far we have made very few assumptions about the general form that the representation space $H$ takes. Now let us take $H$ to be the simplest possible graded space, with just two non-vanishing components which we can call $H_{R}$ and $H_{L}$ (for "right" and "left"): 
\begin{equation}
  H=H_{R}\oplus H_{L}.
\end{equation}
Without loss of generality, we can take $H_{R}$ to be the component of lower grading while $H_{L}$ is the component of higher grading.  Note that, when $\epsilon''=+1$, we should take $H_{R}$ and $H_{L}$ to have ordinary integer gradings ($0$ and $1$, respectively, like the two states in a fermionic fock space), but when $\epsilon''=-1$, we should take $H_{R}$ and $H_{L}$ to have half-integer gradings ($-1/2$ and $+1/2$, respectively, like the two states of a spin-$1/2$ particle).

As explained in Subsection \ref{representing_dga}, since $B$ is a DGA, its differential $d$ must map $m$-forms in $H$ to $(m+1)$-forms in $H$ and so, in particular, must map $H_{R}$ to $H_{L}$, and $H_{L}$ to $0$.  Thus, for $h\in H$, we can write 
\begin{equation}
  d(h)=d_{H}^{}h
\end{equation}
where $d_{H}$ is a linear operator on $H$ that, in the $\{H_{L},H_{R}\}$ basis, is only non-zero in its upper off-diagonal block:
\begin{equation}
  \label{d_H}
  d_{H}=\left(\begin{array}{cc} 0 & \Delta \\ 0 & 0 \end{array}\right).
\end{equation}
In the previous section, we saw that $d^{2}$ vanishes on $A$; and now, since $H$ has only two components, it immediately follows that $d^{2}$ also vanishes on $H$, and hence on the whole of $B$ (as is required in order for $B$ to be a DGA).

So far we have specified $A$ and $H$; but to specify the algebra $B=A\oplus H$, we must also specify the product between $A$ and $H$.  Actually, we only need to specify the product $\hat{a}h$ -- {\it i.e.}\ the left-action of a zero-form $\hat{a}\in \hat{A}$ on an element $h\in H$ -- since (as we shall see) the remaining products $a_{m}h$ and $ha_{m}$ (the left-action or right-action of an arbitrary $m$-form $a_{m}\in A_{m}$ on an element $h\in H$) are then determined by the general structure of an Eilenberg $\ast$-DGA.  To explain this point clearly, let us proceed in three steps: (i) first we discuss the left- and right-action of $\hat{a}$ on $h$; (ii) second we describe the left- and right-action of $d(\hat{a})$ on $h$; and (iii) third we discuss the left- and right-action of $d(\hat{a})^{\ast}$ on $h$.  (Any more complicated element of $A$ is just a product of $\hat{a}$'s, $d(\hat{a})$'s and $d(\hat{a})^{\ast}$'s.)

(i) First we consider the left- and right-action of $\hat{a}$ on $H$:
\begin{subequations}
  \begin{eqnarray}
    L_{\hat{a}}h&=&\hat{a}h, \\
    R_{\hat{a}}h&=&h\hat{a}.
  \end{eqnarray}
\end{subequations}
[Note that, in the NCG literature, the operators $L_{\hat{a}}$ and $R_{\hat{a}}$ (the left and right representations of $\hat{a}$ on $H$) are sometimes denoted $\pi(\hat{a})$ and $\pi(\hat{a})^{0}$, respectively.]  As explained in Subsection \ref{representing_star}, the fact that $B$ is a $\ast$-algebra then implies that $H$ is equipped with an invertible anti-linear operator $J$ ($h^{\ast}=Jh$) and, moreover, that the right-representation $R_{\hat{a}}$ is determined by the left-representation $L_{\hat{a}}$ via the familiar NCG formula
\begin{equation}
  \label{R_hat_from_L_hat}
  R_{\hat{a}}=J L_{\hat{a}^{\ast}} J^{-1}.
\end{equation}

(ii) Next we consider the left- and right-action of $d(\hat{a})$ on $H$: 
\begin{subequations}
  \begin{eqnarray}
    L_{d(\hat{a})}h&=&d(\hat{a})h, \\
    R_{d(\hat{a})}h&=&h d(\hat{a}).
  \end{eqnarray}
\end{subequations}
Since $B$ is a DGA, we can use the Leibniz rule (\ref{graded_Leibniz_Eilenberg}) to determine $L_{d(\hat{a})}$ and $R_{d(\hat{a})}$ in terms of $L_{\hat{a}}$ and $R_{\hat{a}}$:
\begin{subequations}
  \label{LR_da}
  \begin{eqnarray}
    \label{L_da}
    L_{d(\hat{a})}&=&[d_{H},L_{\hat{a}}], \\
    \label{R_da}
    R_{d(\hat{a})}&=&[d_{H},R_{\hat{a}}](-1)^{-|h|},\label{Right_action_da}
  \end{eqnarray}
\end{subequations}
where by $|h|$ we mean the order of the Hilbert space element $h$, according to the grading: {\it i.e.}\ $|h|$ is $0$ or $1$ when $\epsilon''=+1$, and $|h|$ is $\pm1/2$ when $\epsilon''=-1$. Note that $(-1)^{|h|}$ is an operator, not just a number, so care must be taken with regards to its position in Eq.~\eqref{Right_action_da}.

(iii) Finally we consider the left- and right-action of $d(\hat{a})^{\ast}$ on $H$: 
\begin{subequations}
  \label{LR_da_star}
  \begin{eqnarray}
    \label{L_da_star}
    L_{d(\hat{a})^{\ast}}h&=&d(\hat{a})^{\ast}h, \\
    \label{R_da_star}
    R_{d(\hat{a})^{\ast}}h&=&h\,d(\hat{a})^{\ast}.
  \end{eqnarray}
\end{subequations}
Since $B$ is a $\ast$-algebra, we can use Eq.~(\ref{RL_relations}) to determine $L_{d(\hat{a})^{\ast}}$ and $R_{d(\hat{a})^{\ast}}$ in terms of $L_{d(\hat{a})}$ and $R_{d(\hat{a})}$ and hence -- using (\ref{LR_da}) -- in terms of $L_{\hat{a}}$ and $R_{\hat{a}}$
\begin{subequations}
  \begin{eqnarray}
    L_{d(\hat{a})^{\ast}}&=&J R_{d(\hat{a})} J^{-1} = J[d_{H},R_{\hat{a}}](-1)^{-|h|}J^{-1}, \\
    R_{d(\hat{a})^{\ast}}&=&J L_{d(\hat{a})} J^{-1}= J[d_{H},L_{\hat{a}}]J^{-1},
  \end{eqnarray}
\end{subequations}
where once again we stress that the factor $(-1)^{-|h|}$ is an operator on $H$ which does not necessarily commute with $J$ or $d_H$.

We thus see how the left- and right-actions of $\hat{a}$, $d(\hat{a})$ and $d(\hat{a})^{\ast}$ on $H$ are all determined in terms of $L_{\hat{a}}$, the left-representation of $\hat{a}$ on $h$ (and hence how the left- and right-action of {\it any} element of $A$ is determined in terms of $L_{\hat{a}}$).

Let us end this subsection with two remarks:
\begin{itemize}
\item First note that $L_{a^{\ast}}=L_{a}^{\dagger}$ and $R_{a^{\ast}}=R_{a}^{\dagger}$ for zero forms;  but $L_{d(a)^{\ast}}\neq L_{d(a)}^{\dagger}$ and $R_{d(a)^{\ast}}\neq R_{d(a)}^{\dagger}$ for one forms.
\item Second note that, in the $\{H_{L},H_{R}\}$ basis, the quantities $L_{d(a)}$ and $R_{d(a)}$ always have the form:
\begin{equation}
  \label{LR_da_blocks}
  L_{d(a)}\sim R_{d(a)} \sim \left(\begin{array}{cc} 0 & x \\ 0 & 0 \end{array}\right)
\end{equation}
while the quantities $L_{d(a)^{\ast}}$ and $R_{d(a)^{\ast}}$ have a form that depends on $\epsilon''$:
\begin{equation}
  \label{LR_da_star_blocks}
  L_{d(a)^{\ast}}\sim R_{d(a)^{\ast}}\sim\left\{
  \begin{array}{cc}
  \left(\begin{array}{cc} 0 & x \\ 0 & 0 \end{array}\right) & \qquad(\epsilon''=+1) \\
  \left(\begin{array}{cc} 0 & 0 \\ x & 0 \end{array}\right) & \qquad(\epsilon''=-1)
  \end{array}\right.
\end{equation}
\end{itemize}
This will be important below.

\subsection{Comparison with traditional NCG formalism}
\label{NCG_comparison}

Now that we have introduced the algebra $B$ (including the simple two-component structure for $H$ assumed in the previous subsection), we would like to see which aspects of the traditional NCG formalism follow from the requirement that $B$ is an associative Eilenberg $\ast$-DGA.

To facilitate the comparison, it will be convenient to introduce two important operators on $H$ that we have not discussed yet : $D$ and $\gamma$.

(i) Introducing $D$.  From the operator $d_{H}$ (which is not hermitian), we can construct another operator
\begin{equation}
  \label{def_D}
  D\equiv d_{H}+d_{H}^{\dagger}=\left(\begin{array}{cc} 0 & \Delta \\ \Delta^{\dagger} & 0 \end{array}\right)
\end{equation}
that {\it is} Hermitian: $D^{\dagger}=D$.  $D$ is the generalized Dirac operator that appears in the traditional NCG spectral triple $\{\hat{A},H,D\}$. 

(ii) Introducing $\gamma$.  Since $H$ is graded, with two parts, we are free to define a corresponding operator on $H$ which detects this grading.  To match the usual NCG notation, we do this by defining an operator $\gamma$ that equals $-1$ on $H_{L}$ and $+1$ on $H_{R}$; so it is block diagonal in the $\{H_{L},H_{R}\}$ basis, and given by
\begin{equation}
  \label{def_gamma}
  \gamma=\left(\begin{array}{cc} -1 & 0 \\ 0 & +1 \end{array}\right).
\end{equation}

Now, let us see which of the traditional NCG axioms and assumptions follow from the fact that $B$ is {\it associative}.  Associativity of $B$ is equivalent to the requirement that the associator $[b,b',b'']$ vanishes, for all $b,b',b''\in B$.  Thus, let us list all the different types of associators that arise in this way, and the consequences of requiring them to vanish.  
\begin{itemize}
\item From $[\hat{a},\hat{a}',\hat{a}'']$ we recover the usual assumption that the coordinate algebra $\hat{A}$ ({\it i.e.}\ the algebra that appears in the traditional spectral triple $\{\hat{A},H,D\}$) is associative;
\item From $[a,a',a'']$ we recover the usual assumption that $A$, the universal algebra of forms over $\hat{A}$, is associative.
\item From $[\hat{a},\hat{a}',h]$ we recover $L_{\hat{a}}L_{\hat{a}'}=L_{\hat{a}\hat{a}'}$ or, in other words, $\pi(\hat{a})\pi(\hat{a}')=\pi(\hat{a}\hat{a}')$, which is the familiar condition satisfied by the left-representation of $\hat{A}$.
\item From $[h,\hat{a}',\hat{a}]$ we recover $R_{\hat{a}}R_{\hat{a}'}=R_{\hat{a}'\hat{a}}$ or, in other words, $\pi(\hat{a})^{0}\pi(\hat{a}')^{0}=\pi(\hat{a}'\hat{a})^{0}$, which is the familiar condition satisfied by the right-representation of $\hat{A}$.
\item From $[\hat{a},h,\hat{a}']$ we recover $[L_{\hat{a}},R_{\hat{a}'}]=0$ or, in other words, $[\pi(\hat{a}),\pi(\hat{a}')^{0}]=0$, which is the traditional "order zero" axiom of NCG.
\item From $[d(\hat{a}),h,\hat{a}']$, $[\hat{a},h,d(\hat{a}')]$, $[d(\hat{a})^{\ast},h,\hat{a}']$ and $[\hat{a},h,d(\hat{a}')^{\ast}]$ we recover 
$[[D,\pi(\hat{a})],\pi(\hat{a}')^{0}]=0$, which is the familiar "order-one" axiom of NCG.
\item All further associators, including those involving two or more $h$'s, or two or more $d(a)$'s or $d(a)^{\ast}$'s are automatically  satisfied and do not yield further constraints, {\it except}:
\item $[d(\hat{a}),h,d(\hat{a}')^{\ast}]$ and $[d(\hat{a})^{\ast},h,d(\hat{a}')]$ which {\it only yield non-trivial constraints when $\epsilon''=-1$}, as may be seen from Eqs.~(\ref{LR_da_blocks}, \ref{LR_da_star_blocks}).  

Note that, unlike the previous associators, which all corresponded to traditional NCG axioms and assumptions, these final two associators correspond to {\it new} "second-order conditions" which are not normally imposed as axioms in NCG.  These new second-order conditions were first pointed out in \cite{Boyle:2014wba}; but the fact that they are only non-trivial when $\epsilon''=-1$ is one of the most important new results in the present paper, and is important for resolving a key puzzle in the NCG construction of the standard model of particle physics.  (We will return to this point in the following section.)
\end{itemize}
Next, as explained in Subsection \ref{representing_star}, from the assumption that $B$ is a $\ast$-algebra we recover:
\begin{itemize}
\item the assumption that $\hat{A}$ and $A$ are $\ast$-algebras;
\item the assumption that $H$ is equipped with an invertible anti-linear operator $J$;
\item the assumption that $J^{2}=\epsilon$ with $\epsilon=\pm1$; and
\item the assumption that $\hat{A}$ is not just left-represented or right-represented on $H$, but rather left-right bi-represented on $H$, with 
the right representation related to the left representation by the familiar NCG formula (\ref{R_hat_from_L_hat}).
\end{itemize}
Finally, let us see what follows from the fact that $B$ is $\ast$-DGA ({\it i.e.}\ from including the differential-graded structure of $B$):
\begin{itemize}
\item From Eqs.~(\ref{def_D}, \ref{def_gamma}), we recover the assumption that $\{D,\gamma\}=0$ (which, from our new perspective, follows from the fact that $d$ is an order-one operator on $H$).
\item The fact that $B$ is graded implies that the zero form $\hat{a}\in\hat{A}$ must map $H_{L}\to H_{L}$ and $H_{R}\to H_{R}$ which implies that in the $\{H_{L},H_{R}\}$ basis, $L_{a_{0}^{}}$ and $R_{a_{0}^{}}$ are block-diagonal, from which we recover the usual NCG axioms 
$[\gamma,L_{\hat{a}}]=[\gamma,R_{\hat{a}}]=0$ or, in other words, $[\gamma,\pi(\hat{a})]=[\gamma,\pi(\hat{a})^{0}]=0$.  
\item From the fact that, as explained in point (v)$'$ of Section \ref{dga} the $\ast$-operation must map $m$-forms to $(\epsilon''m)$-forms, we recover the assumption that $J\gamma=\epsilon''\gamma J$ where $\epsilon''=\pm1$ is the $\ast$-DGA sign choice explained in Section \ref{dga}.
\item Finally, as explained in Subsection \ref{defining_A}, to make $A$ a $\ast$-DGA, we require $d(d(a)^{\ast})=0$ and hence 
\begin{equation}
  \label{new_2nd_order}
  L_{d(d(\hat{a})^{\ast})}=\{d_{H},L_{d(\hat{a})^{\ast}}\}=0.
\end{equation}  
Note that, unlike the previous few conditions which all corresponded to traditional NCG axioms and assumptions, this final condition corresponds to a {\it new} condition (like the new second-order condition described above); and (again like the second order condition) this new condition is only non-trivial when $\epsilon=-1$.
\end{itemize}

So far in this subsection, we have focused on explaining how the requirement that $B$ is an associative $\ast$-DGA, together with the simplest choice of grading on $H$, unifies and explains a long list of the traditional axioms and assumptions of NCG.  Let us end this Subsection by summarizing a few of the key ways in which our construction {\it differs} from traditional NCG:
\begin{itemize}
\item We differ from the traditional NCG formalism in our representation of how $d(\hat{a})$ and $d(\hat{a})^{\ast}$ act on $H$.  Fundamentally this is because our grading for $H$ differs from the traditional one in NCG.
\item We obtain novel "second-order" constraints $[d(\hat{a}),h,d(\hat{a}')^{\ast}]=[d(\hat{a})^{\ast},h,d(\hat{a}')]=0$ which are only non-trivial when $\epsilon''=-1$.
\item We obtain another novel constraint (\ref{new_2nd_order}) which is only non-trivial when $\epsilon''=-1$.
\end{itemize}
In the following section, we will see how these differences resolve several key puzzles in the NCG formulation of the standard model of particle physics.

\section{Application to the Standard Model}
\label{standard_model}

In the previous section, we defined a particularly simple type of Eilenberg algebra $B=A\oplus H$, where $A$ was the universal $\ast$-algebra of differential forms over some "algebra of coordinates" $\hat{A}$, and $H$ was the simplest non-trivially graded space (with just two components, $H_{R}$ and $H_{L}$); and we explained how the traditional list of NCG axioms and assumptions could, to a remarkable degree, be unified in the requirement that $B$ was an associative $\ast$-DGA.  In order to specify a {\it particular} $B$ of this type, it just remains to choose the following three final inputs: (i) a particular algebra of coordinates $\hat{A}$; (ii) a particular left-representation of $\hat{A}$ on $H$; and (iii) particular values for the $\pm$ signs $\epsilon$ and $\epsilon''$.  

In this section, we choose these final inputs to be the ones used in the traditional NCG description of the standard model, and then investigate the consequences.    In the traditional approach, the geometry of the standard model is, itself, the product of two geometries: (i) one that describes an ordinary, continuous, commutative four-dimensional spacetime; and (ii) another that describes a finite, discrete, non-commutative "internal space".  In our language, this corresponds to an Eilenberg algebra $B$ that is, itself, the product of two Eilenberg algebras $B_{c}$ and $B_{f}$.  We discuss these two algebras in turn -- for each, we begin by listing the three final inputs mentioned in the previous paragraph.

For a pedagogical introduction to the traditional NCG construction of the standard model, see \cite{vandenDungen:2012ky, vanSuijlekomBook}.  Note that in this section we discuss the {\it geometry} of the standard model.  Traditionally one then uses the so-called ``spectral action" formula to convert this geometry into an ordinary field theory Lagrangian (which, for this particular choice of geometry, turns out to be the Lagrangian for the standard model of particle physics coupled to Einstein gravity).  For an introduction to this conversion process (which we do not discuss here), see \cite{vandenDungen:2012ky}.  Also note that, in what follows, it will be convenient to discuss the continuous and finite parts of the geometry separately.
The bosonic fields of the theory (in particular, the gauge fields and Higgs fields) then arise from the interaction between these two parts (again see \cite{vandenDungen:2012ky} for an introduction).

\subsection{The continuous geometry, $B_{c}$}  
\label{continuous_geometry}

The continuous geometry is described by the so-called "canonical spectral triple" corresponding to ordinary (commutative) four-dimensional Riemannian geometry (see Section 2.1 in \cite{vandenDungen:2012ky}).

(i) In this case, the algebra of coordinates is $\hat{A}=C^{\infty}(M)$, the algebra of smooth functions on the compact 4-dimensional Riemannian spin manifold $M$.

(ii) $H=H_{R}\oplus H_{L}$ is the Hilbert space of square-integrable (Dirac) spinors  on $M$.  $H_{R}$ and $H_{L}$ are the subspaces of right-handed and left-handed Weyl spinors.  $\hat{A}$ is represented on $H$ by pointwise multiplication: $(ah)(x)=a(x)h(x)$ for $a(x)\in\hat{A}$ and $h(x)\in H$.

(iii) In this case, the signs $\epsilon$ and $\epsilon''$ are fixed by the representation theory of Clifford algebras; the appropriate values depend on the dimension (mod 8) of the manifold $M$ (see {\it e.g.}\ the table in Section 2.2.2 of \cite{vandenDungen:2012ky}).  Since the traditional NCG description of the standard model is formulated in Euclidean signature,\footnote{See the Discussion for a related comment.} the values in 4D are $\epsilon=-1$ and $\epsilon''=+1$.   

In this case, the operator $D$ (\ref{def_D}) is just the ordinary curved-space Dirac operator
\begin{equation}
  D=D\!\!\!\!/\;\;=-i\gamma^{\mu}\nabla_{\mu}^{S},
\end{equation}
where $\nabla_{\mu}^{S}$ is the Levi-Civita spin connection and the $\gamma^{\mu}$ are the gamma matrices.  

In the previous section we showed that our formalism predicts new geometrical constraints, in addition to those of the traditional NCG formalism -- but these new constraints are only non-trivial when $\epsilon''=-1$.  As a result, since the continuous geometry described above has $\epsilon''=+1$, it only needs to satisfy the traditional NCG constraints (which it does), and it does {\it not} need to satisfy any additional second-order constraint.  

This neatly resolves a key puzzle that was noted at the end of our earlier paper \cite{Boyle:2014wba} and emphasized in \cite{Brouder:2015qoa}.  In our earlier construction \cite{Boyle:2014wba}, the second-order condition gave a non-trivial constraint on both the finite and continuous parts of the geometry; and, although the constraint on the finite part was successful (in that it neatly removed all of the unwanted terms in the finite Dirac operator), the constraint on the continuous part was too strong.  By constrast, here we see that the second-order condition only gives a non-trivial constraint in the $\epsilon=-1$ case (so it still gives a successful non-trivial constraint on the finite geometry, which has $\epsilon''=-1$, but it gives no additional constraint on the continuous geometry, which has $\epsilon''=+1$, and hence is perfectly compatible with it).

\subsection{The finite geometry, $B_{f}$}  
\label{finite_geometry}

The discrete "internal space" is described by a finite spectral triple (with a finite-dimensional algebra, represented on a finite-dimensional Hilbert space) -- see Section 6 in \cite{vandenDungen:2012ky}.

(i) In this case, the algebra of coordinates $\hat{A}$ is the direct sum of the complex numbers $\mathbb{C}$, the quaternions $\mathbb{H}$ and the $3\times3$ complex matrices $M_{3}(\mathbb{C})$: 
\begin{equation}
  \hat{A}=\mathbb{C}\oplus\mathbb{H}\oplus M_{3}(\mathbb{C}).
\end{equation}

(ii) Next, we describe the left-representation of $\hat{A}$ on $H$.  For clarity, we will describe a single generation of standard model fermions; the extension to the full set of three generations is straightforward.  The Hilbert space $H$ is $\mathbb{C}^{32}$ (here 32 is the number of fermionic degrees of freedom in a single standard model generation, if we include a right-handed neutrino in each generation to account for the observed neutrino masses).  The left-representation of $\hat{A}$ on $H$ is block-diagonal: let us start by giving physically-appropriate names to the corresponding subspaces of $H$ on which these blocks act.  For starters, as explained in Subsection \ref{defining_B}, the grading splits $\mathbb{C}^{32}$ into two copies of $\mathbb{C}^{16}$: $H=H_{L}\oplus H_{R}$.  Next, each $\mathbb{C}^{16}$ splits into two copies of $\mathbb{C}^{8}$: $H_{L}=\bar{F}_{R}\oplus F_{L}$ and $H_{R}=F_{R}\oplus\bar{F}_{L}$ (here the subspaces $F_{L}$ and $F_{R}$ contain the left- and right-handed fermions, while $\bar{F}_{L}$ and $\bar{F}_{R}$ contain the corresponding anti-fermions).  Finally, each $\mathbb{C}^{8}$ splits into a lepton part ($\mathbb{C}^{2}$) and a quark part ($\mathbb{C}^{2}\otimes\mathbb{C}^{3}$):
\begin{eqnarray}
  F_{L}={\cal L}_{L}\oplus{\cal Q}_{L} 
  &\qquad&
  F_{R}={\cal L}_{R}\oplus{\cal Q}_{R} \nonumber \\
  \bar{F}_{L}=\bar{{\cal L}}_{L}\oplus\bar{{\cal Q}}_{L} 
  &\qquad&
  \bar{F}_{R}=\bar{{\cal L}}_{R}\oplus\bar{{\cal Q}}_{R}.
\end{eqnarray}
Now, if we consider an arbitrary element $\hat{a}=(\lambda,q,m)\in \hat{A}$, where $\lambda\in\mathbb{C}$ is a complex number, $q\in\mathbb{H}$ is a quaternion, and $m\in M_{3}(\mathbb{C})$ is a $3\times3$ complex matrix, and we write
\begin{equation}
  \label{q}
  q=\left(\begin{array}{cc} \alpha & \beta \\ -\bar{\beta} & \bar{\alpha} \end{array}\right)\qquad{\rm and}\qquad
  q_{\lambda}^{}=\left(\begin{array}{cc} \lambda & 0 \\ 0 & \bar{\lambda} \end{array}\right)
\end{equation}
where $\alpha$ and $\beta$ are complex numbers, then (in the $\{\bar{{\cal L}}_{R},\bar{{\cal Q}}_{R},{\cal L}_{L},{\cal Q}_{L},{\cal L}_{R},{\cal Q}_{R},\bar{{\cal L}}_{L},\bar{{\cal Q}}_{L}\}$ basis just described) the left-action of $\hat{a}$ on $H$ has the following block diagonal form
\begin{equation}
  L_{\hat{a}}=\left(\begin{array}{cccc|cccc}
  \lambda\mathbb{I}_{2} &&&&&&& \\
  &\mathbb{I}_{2}\otimes m &&&&&& \\
  && q &&&&& \\
  &&& q\otimes\mathbb{I}_{3} &&&& \\
  \hline
  &&&& q_{\lambda}^{} &&& \\
  &&&&& q_{\lambda}^{}\otimes\mathbb{I}_{3} && \\
  &&&&&& \lambda\mathbb{I}_{2} & \;\qquad\; \\
  \;\qquad\;&\;\qquad\;&\;\qquad\;&\;\qquad\;&\;\qquad\;&\;\qquad\;&\;\qquad\;& \mathbb{I}_{2}\otimes m \end{array}\right),
\end{equation}
where $\mathbb{I}_{2}$ and $\mathbb{I}_{3}$ denote the $2\times2$ and $3\times3$ identity matrices, respectively.

(iii) In this case, the space has KO dimension 6: the corresponding signs are $\epsilon=+1$ and $\epsilon''=-1$ (along with $DJ=JD$,  again see the table in Section 2.2.2 of \cite{vandenDungen:2012ky}).  In the basis just described, $J$ and $\gamma$ are then given by
\begin{equation}
  \label{J_gamma}
  J=\left(\begin{array}{cc} 0 & \;\mathbb{I}\; \\ \;\mathbb{I}\; & 0 \end{array}\right)\circ {\rm c.c.}
  \qquad{\rm and}\qquad
  \gamma=\left(\begin{array}{cc} -\mathbb{I} & 0 \\ 0 & +\mathbb{I} \end{array}\right)
\end{equation}
where $\mathbb{I}$ is the $16\times16$ identity matrix, and "c.c." stands for "complex conjugation" -- a reminder that $J$ is anti-linear.  

From the traditional NCG conditions $D_{F}=D_{F}^{\dagger}$, $\{D_{F},\gamma_{F}\}=0$, $[D_{F},J_{F}]=0$ and the 
order one condition, one finds that $D_{F}$ is constrained to the form
\begin{equation}
  \label{D_F}
  D_{F}=\left(\begin{array}{cc} 0 & \Delta \\
  \Delta^{\dagger} & 0 \end{array}\right), 
\end{equation}
where $\Delta$ is a $16\times16$ symmetric matrix of the form
\begin{equation}
  \label{Delta}
  \Delta=\left(\begin{array}{cc|cc} 
  M & N^{T} & Y_{l}^{T} & 0 \\
  N & 0 & 0 & Y_{q}^{T} \\
  \hline
  Y_{l}^{} & 0 & 0 & 0 \\
  0 & Y_{q}^{} & 0 & 0 \end{array}\right).
\end{equation}
Here $Y_{l}$ and $Y_{q}$ are two arbitrary $2\times2$ matrices, $M$ is a $2\times2$ symmetric matrix given by
\begin{equation}
  \label{M}
  M=\left(\begin{array}{cc} a & b \\ b & 0 \end{array}\right)
\end{equation}
where $a$ and $b$ are arbitary constants, and $N$ is a $6\times2$ matrix given by
\begin{equation}
  \label{N}
  N=\left(\begin{array}{cc} \vec{c} & \vec{d} \\ 0 & 0 \end{array}\right),
\end{equation}
where $\vec{c}$ and $\vec{d}$ are two arbitrary $3\times1$ column vectors.  

As reviewed in Ref.~\cite{Boyle:2014wba}, these traditional NCG constraints are not strong enough -- {\it i.e.}\ they do not constrain $D$ to be of the phenomenologically desired form.  Instead, they allow non-zero values for the parameters $b$, $\vec{c}$ and $\vec{d}$, which give rise to phenomenologically unwanted Yukawa couplings and scalar fields if they are not eliminated.  This problem was highlighted in \cite{Chamseddine:2006ep} and the concluding section of \cite{Chamseddine:2007hz}.  In the traditional formalism, one is forced to introduce an extra (empirically-motivated, non-geometric) condition (called the "massless photon" condition \cite{Chamseddine:2006ep, Chamseddine:2007hz}) in order to eliminate these extra, unwanted terms.   

In our formalism, something nice happens instead.  As explained in Subection \ref{NCG_comparison}, when $\epsilon''=-1$ (as is the case for the finite geometry), we obtain non-trivial "second-order conditions" $[d(\hat{a}),h,d(\hat{a}')^{\ast}] =[d(\hat{a})^{\ast},h,d(\hat{a}')]=0$.  These second-order conditions yield exactly the same restrictions on $D_{F}$ that we previously obtained from the second-order conditions in Ref.~\cite{Boyle:2014wba} (or that Brouder {\it et al} later obtained from the second-order conditions in \cite{Brouder:2015qoa}).  In particular, as shown in \cite{Boyle:2014wba}, these second-order constraints may be satisfied in four different ways by setting (i) $b=\vec{c}=\vec{d}=0$;  (ii) $Y_{q,11}^{}=Y_{q,21}^{}=b=0$; (iii) $Y_{l,11}^{}=Y_{l,21}^{}=\vec{c}=\vec{d}=0$; or (iv) $Y_{l,11}^{}=Y_{l,21}^{}=Y_{q,11}=Y_{q,21}=\vec{c}=0$.   In particular, solution (i) precisely corresponds to setting the seven unwanted coeffients ($b$, $\vec{c}$, $\vec{d}$) to zero, without having to introduce the extra non-geometrical massless photon condition. 

Thus, our current formalism neatly resolves the paradox in our earlier paper \cite{Boyle:2014wba}: that our new second-order constraint seemed to provide such a phenomenologically successful constraint on the finite part of the geometry, while at the same time providing an unwanted over-constraint on the continuous part of the geometry.   In our new formalism, the successful constraint on the finite part of the geometry is retained, while the overconstraint on the continuous part is eliminated.  This resolution is satisfying, as it directly follows from thinking more clearly about the basic structure of a $\ast$-DGA -- see Remark (v)' in Section \ref{dga}.

\subsection{Applying the new constraint}
\label{new_constraints}

In the previous two subsections, we have explained how our present formalism, when applied to the standard model ``input data" (the choice of $\hat{A}$ and its representation on $H$ described above), recovers all the successful constraints from our earlier paper \cite{Boyle:2014wba} (including the successful second-order condition on the finite geometry in \cite{Boyle:2014wba}), while also neatly avoiding the problematic second-order {\it over}-constraint on the continuous geometry in \cite{Boyle:2014wba}.

Now we turn to our present construction's {\it new} constraint (\ref{new_2nd_order}), which applies when $\epsilon''=-1$ (as is the case for the finite geometry).  This constraint is new, not only compared to traditional (spectral triple) NCG formalism, but also compared to our formalism in \cite{Boyle:2014wba}.

First, it is worth noting that, if we look back at the construction of $A$ from $\hat{A}$ in Subsection \ref{defining_A}, we see that the new constraint (\ref{new_2nd_order}) has a different status from the other constraints that we have discussed.  In particular, it is not actually needed in the definition or construction of the graded $\ast$-algebras $A$ and $B$.\footnote{The graded $\ast$-algebra $A$ (consisting of arbitrary products of $\hat{a}$'s, $d(\hat{a})$'s and $d(\hat{a})^{\ast}$'s, and arbitrary linear combinations of such products) is obtained by first defining a first-order differential calculus ($A_{1},d$) over $\hat{A}$ \cite{DuboisViolette:1999cj}, and then tensoring this (over $\hat{A}$) with itself arbitrarily many times.}  Instead, it is only needed when we formally extend the differential $d$ from $\hat{A}$ to $A$, so that we can reinterpret $A$ and $B$ as not merely graded $\ast$-algebras, but as {\it differential} graded $\ast$-algebras, and pass from the level of first-order differential calculus to full differential calculus \cite{DuboisViolette:1999cj}.

If we now impose the complete set of constraints -- {\it i.e.}\ those discussed in Subsection \ref{finite_geometry} {\it plus} the new constraint (\ref{new_2nd_order}) -- we find that the following striking results:
\begin{itemize}
\item (i) On the one hand, if we take the {\it full} finite algebra $\hat{A}=\mathbb{C}\oplus\mathbb{H}\oplus M_{3}(\mathbb{C})$, with the representation described in Subsection \ref{finite_geometry}, the complete set of constraints can only be simultaneously satisfied if the finite Dirac operator vanishes identically: $D_{F}=0$.
\item (ii) On the other hand, if we restrict the full finite algebra $\hat{A}$ (and its representation on $H$) to an appropriate sub-algebra $\hat{A}'$, then the complete set of constraints can be satisfied for {\it non}-vanishing $D_{F}$.  In particular, one can check that the {\it maximal} sub-algebra compatible with non-vanishing $D_{F}$ is $\hat{A}'=\mathbb{C}\oplus M_{3}(\mathbb{C})$.  When we restrict the full algebra $\hat{A}$ to the sub-algebra $\hat{A}'$, the corresponding representation of an element $\{\lambda,m\}\in \hat{A}'$ is obtained by restricting the representation of an element $\{\lambda,q,m\}\in\hat{A}$ (described in Subsection \ref{finite_geometry}) as follows: wherever the $2\times2$ quaternion $q$ appears, we make the replacement:
\begin{equation}
  q=\left(\begin{array}{cc} \alpha & \beta \\ -\bar{\beta} & \bar{\alpha} \end{array}\right)
  \qquad\rightarrow\qquad
  q_{\lambda}(\hat{n})=
  {\rm Re}(\lambda)I+i\,{\rm Im}(\lambda)\hat{n}\cdot\vec{\sigma},
\end{equation}
where $I$ is the $2\times2$ identity matrix, $\vec{\sigma}=\{\sigma_{1},\sigma_{2},\sigma_{3}\}$ are the three Pauli sigma matrices, and $\hat{n}=\{{\rm sin}(\theta){\rm cos}(\varphi), {\rm sin}(\theta){\rm sin}(\varphi),{\rm cos}(\theta)\}$ is a unit vector in $\mathbb{R}^{3}$.  In other words, $q_{\lambda}(\hat{n})$ represents the general embedding of $\mathbb{C}$ in $\mathbb{H}$ (the unit vector $\hat{n}$ describes the orientation of the single complex imaginary $i\in\mathbb{C}$ in the 3D space of quaternionic imaginaries $\{I,J,K\}$); and the previous diagonal embedding $q_{\lambda}$ defined in Eq.~(\ref{q}) corresponds to the special case $\hat{n}=\{0,0,1\}$.

Now if we restrict to this sub-algebra and its representation, and apply the complete set of constraints ({\it i.e.}\ if we use $\hat{A}'$ and its representation on $H$ to build the corresponding algebra $B'$, and then demand that it is a $\ast$-DGA), we find that $D_{F}$ is again restricted to have the form given in Eqs.~(\ref{D_F}, \ref{Delta}, \ref{M}, \ref{N}); but now, in addition, we find that the parameters $b$, $\vec{c}$, and $\vec{d}$ in $D_{F}$ must vanish
\begin{equation}
  b=\vec{c}=\vec{d}=0,
\end{equation}
while the Yukawa matrices $Y_{l}$ and $Y_{q}$ in $D_{F}$ are restricted to the form (note the sign change compared to the convention we used in \cite{Boyle:2014wba}):
\begin{equation}
  Y_{l}=\left(\begin{array}{cc} 
  \;+y_{\nu}^{}\varphi_{1}^{}\; & \;\;\;y_{e}^{}\bar{\varphi}_{2}^{}\; \\
  \;-y_{\nu}^{}\varphi_{2}^{}\; & \;\;\;y_{e}^{}\bar{\varphi}_{1}^{}\; \end{array}\right),
  \qquad\qquad
  Y_{q}=\left(\begin{array}{cc} 
  \;+y_{u}^{}\varphi_{1}^{}\; & \;\;\;y_{d}^{}\bar{\varphi}_{2}^{}\; \\
  \;-y_{u}^{}\varphi_{2}^{}\; & \;\;\;y_{d}^{}\bar{\varphi}_{1}^{}\; \end{array}\right),\label{Yukawa_Couple}
\end{equation}
where the four complex numbers $\{y_{\nu},y_{e},y_{u},y_{d}\}$ are arbitrary, while the two complex numbers $\{\varphi_{1},\varphi_{2}\}$ are constrained to satisfy:
\begin{equation}
  \label{Phi_eq}
  (I+\hat{n}\cdot\vec{\sigma})\left(\begin{array}{c} \bar{\varphi}_{2} \\ \bar{\varphi}_{1} \end{array}\right)=0.
\end{equation}

$D_{F}$ now has {\it precisely} the correct form to match the standard model of particle physics (see {\it e.g.}\ \cite{vandenDungen:2012ky} for a more detailed explanation): the four complex numbers $\{y_{\nu},y_{e},y_{u},y_{d}\}$ become the neutrino, electron, up quark and down quark Yukawa couplings, respectively, and $\Phi=\{\varphi_{1},\varphi_{2}\}$ becomes the usual standard model Higgs doublet.\footnote{More precisely, $\Phi$ will correspond to the Higgs doublet upon fluctuation of the Dirac operator in the full product geometry (again see {\it e.g.}\ \cite{vandenDungen:2012ky} for a more detailed explanation).}  Note that the {\it same} Higgs doublet $\Phi$ appears in both $Y_{l}$ and $Y_{q}$; and if we include all three generations, $Y_{l}$ and $Y_{q}$ still have the form shown in Eq.~\eqref{Yukawa_Couple}: $y_{\nu}$, $y_{e}$, $y_{u}$ and $y_{d}$ become $3\times3$ Yukawa coupling matrices, but there is still just a single Higgs doublet.

We emphasize that the constraints on $D_{F}$ in our earlier paper \cite{Boyle:2014wba} (and in Subsection \ref{finite_geometry} above) had two shortcomings.  First, our ``second-order condition" was only strong enough to restrict $D_{F}$ to one of four possible forms (and we had to choose the phenomenologically correct form from among these four by hand).  Second, even after we chose the correct form from among these four options, $D_{F}$ was still too general since the matrices $Y_{l}$ and $Y_{q}$ were not yet constrained to the  phenomenologically correct form (\ref{Yukawa_Couple}).  

It is remarkable that the new constraint (\ref{new_2nd_order}) resolves both of these shortcomings (so that $D_{F}$ is restricted to precisely the phenomenologically correct form), while at the same time forcing spontaneous breaking from $\hat{A}$ to $\hat{A}'$, which corresponds to the usual electroweak symmetry breaking of the full standard model gauge group $SU(3)\times SU(2)\times U(1)_{Y}$ down to the unbroken subgroup $SU(3)\times U(1)_{E\& M}$.  

As explained above, the embedding of $\mathbb{C}$ in $\mathbb{H}$ involves choosing an arbitrary direction $\hat{n}$.  This choice may be made independently over each point in the continuous 4D spacetime, and is linked to the ``direction" of the Higgs doublet $\Phi$ at that same point in spacetime via Eq.~(\ref{Phi_eq}).  We see that, in our new formalism, the unbroken $SU(3)\times U(1)$ gauge symmetries of the standard model correspond to the inner derivations of $B'$, while the remaining three broken generators (corresponding to the massive $W^{\pm}$ and $Z$ bosons) act non-trivially on $\Phi$ (and rotate $\hat{n}$).

It is striking that, in our present formalism, electroweak symmetry breaking already appears before we write down the action: it arises when we want to extend the differential $d$ from $\hat{A}$ to $B$, to promote $B$ from a graded $\ast$-algebra to a {\it differential}-graded $\ast$-algebra, and pass from first-order differential calculus to full differential calculus.
\end{itemize}

\section{Discussion}
\label{Discussion}

Let us briefly recap.  We start by (i) choosing an ``algebra of coordinates" $\hat{A}$; (ii) defining $A$, the corresponding universal $\ast$-algebra of differential forms over $A$; and (iii) taking the simplest non-trivially graded representation of $A$ on $H$ ({\it i.e.}\ the one where $H$ has just two non-zero components, $H_{L}$ and $H_{R}$).  As we explain in Section \ref{NCG}, nearly all of the axioms and assumptions of the traditional real-spectral-triple formalism of NCG are then recovered from the simple requirement that the corresponding Eilenberg algebra $B=A\oplus H$ (a particularly simple type of super-algebra) is a $\ast$-DGA.  Moreover, this requirement also implies other, novel, geometric constraints.  As we explain in Section \ref{standard_model}, when we apply our current construction to the specific NCG data traditionally used to describe the standard model of particle physics, we find that these new constraints are physically meaningful and phenomenologically correct.

Our current construction improves on our earlier framework~\cite{Boyle:2014wba, Farnsworth:2014vva} in a number of important respects.  From a mathematical standpoint, our earlier definition of $B$ in \cite{Boyle:2014wba, Farnsworth:2014vva} already unified some of the NCG axioms and assumptions, but now we are able to go much further: roughly speaking, the earlier definition of $B$ unified the axioms and assumptions that did {\it not} involve $\gamma$, while the new definition also incorporates those that {\it do} involve $\gamma$.  It is encouraging that this improved unification goes hand-in-hand with the improved mathematical structure of $B$ which (with its new definition) is now a proper $\ast$-DGA.  From a physical standpoint, our new formalism resolves a paradox which arose in our earlier work~\cite{Boyle:2014wba, Farnsworth:2014vva}: namely, in our earlier formalism, the second-order condition (when applied to the standard model geometry) seemed to give a phenomenologically successful constraint on the finite part of the geometry, but a problematic over-constraint on the continuous part of the geometry. In our current formalism, the unwanted over-constraint on the continuous geometry is automatically eliminated (as explained in Subsection \ref{continuous_geometry}), while the successful constraint on the finite geometry is automatically retained (as explained in Subsection \ref{finite_geometry}).  It is satisfying that this resolution follows from thinking more clearly about the basic structure of a $\ast$-DGA -- see Remark (v)' in Section \ref{dga}.  As explained in Subsection \ref{new_constraints}, the formalism developed in the present paper also predicts a constraint (\ref{new_2nd_order}) that is new, not only compared to traditional (spectral triple) NCG formalism, but also compared to our formalism in \cite{Boyle:2014wba}.  Strikingly, this new constraint turns out to fix $D_{F}$ (the ``Dirac operator" on the finite space) to precisely the phenomenologically desired form while, at the same time, requiring electroweak symmetry breaking, and providing a new geometric interpretation for this basic aspect of the standard model of particle physics.

The construction presented in this paper also inherits a few other nice features from our earlier work: (i) first, the conceptually nice reinterpretation of the symmetries of the standard model, and the structure of the gauge-Higgs sector, as arising from the requirement that the action should be invariant under automorphisms of $B$ (see \cite{Farnsworth:2014vva}); and (ii) second, the corresponding implication that the traditional "standard model geometry" actually yields a slight {\it extension} of the standard model which, in addition to including a right-handed partner for each left-handed neutrino, also includes an extra $U(1)_{B-L}$ gauge symmetry and, correspondingly, two new particles: a new $U(1)_{B-L}$ gauge boson, and a new complex scalar field $\sigma$ that is a singlet under the standard model gauge group $SU(3)\times SU(2)\times U(1)$, but is charged under the new $U(1)_{B-L}$, and is responsible for Higgsing this symmetry (so that it is unseen at low energies).  As emphasized in \cite{Farnsworth:2014vva}, this extension of the standard model is phenomenologically viable, and resolves the discrepancy between the traditional NCG Higgs mass prediction ($\sim170~{\rm GeV}$) with that of the observed Higgs mass ($\sim125~{\rm GeV}$), and can account for several cosmological observations that cannot be accounted for by the standard model alone \cite{Boyle:2011fq}.

It is also worth mentioning that there are still some puzzling and unsatisfying features of the construction presented here.  For example, the standard model geometry as we have described it looks like a real algebra represented on a complex Hilbert space.  Really, we should interpret this as short-hand for a real algebra represented on a real Hilbert space of double the dimension (and this is also what we must do if we want to regard $A$ and $H$ as two subspaces of a common vector space $B=A\oplus H$).  This has an awkward consequence: in Section \ref{NCG}, we argued that the antilinearity of $J$ (over $\mathbb{C}$) could be {\it derived} from the fact that $B$ is a $\ast$-algebra; but if $B$ is a $\ast$-algebra over $\mathbb{R}$, then this is no longer true, and the anti-linearity of $J$ (over $\mathbb{C}$) must be put in by hand.  We regard this awkward and seemingly technical point as a clue that there is something important that is still missing from the formalism presented here, and we are currently working on a follow-up paper addressing this issue among others \cite{Jordan}.

It is natural to wonder about the physical implications of the new geometrical interpretation of electroweak symmetry breaking described above.  In the traditional NCG formulation of the standard model, one uses the "spectral action" which, in order to roughly match observations, must live at roughly the grand unification scale ($\sim10^{16}~{\rm GeV}$), and yields certain relations between coupling constants at that scale (including the standard gauge-coupling relation familiar from grand unification).  It seems that the formalism presented in the present paper should naturally "live" at the electroweak scale, rather than the GUT scale, and so it is interesting to consider whether or how this might be made compatible with the traditional spectral action formula.  Perhaps a new formula for the bosonic action will be needed in this new framework.  We leave these interesting issues for future work.

There are also many other possible directions for future work. 
 (i) Firstly, as noted in \cite{Brouder:2015qoa}, the DGA structure of $B$ suggests a connection to the BRST/BV formalism, and may be an important clue about how to quantize correctly.  (ii) It is interesting to consider how our formalism interacts with other recent interesting proposals for how to go beyond the traditional NCG formulation of the standard model (see {\it e.g.}\ \cite{Chamseddine:2013rta, Chamseddine:2013kza, Chamseddine:2014nxa, Chamseddine:2014uma, Chamseddine:2015ata, Devastato:2013oqa, Devastato:2014bta, Iochum:2003xy, D'Andrea:2014dya, Barrett:2015naa}).  (iii) We believe that there are interesting issues to be sorted out involving the relation between euclidean and lorentzian signature -- see the comment in Subsection \ref{finite_geometry}.  (iv) Although we have mostly restricted our attention in this paper to the case where the algebras $\hat{A}$, $A$ and $B$ are associative, we were originally led to Eilenberg's approach by the fact that it was designed to naturally generalize to the non-associative case \cite{Farnsworth:2013nza, Boyle:2014wba}.  In other words, one of the advantages of our new formalism is that it is naturally suited to generalizing from non-commutative to non-associative geometry; and this continues to seem like a very interesting avenue to explore.  In particular, it is intriguing to explore physical models based on a coordinate algebra $\hat{A}$ involving the octonions, or the exception Jordan algebra.  (v) Finally, in the current paper we assume the simplest possible non-trivial representation space, identifying the usual $Z_2$ grading on $H$ with its differential grading. A natural extension of the ideas discussed here would be to consider more general implementations of the grading on $H$.

\acknowledgments

We would like to thank John Barrett, Fabien Besnard, Nadir Bizi, Christian Brouder, Tobias Fritz, Masoud Khalkhali and Matilde Marcolli for useful discussions during the writing of this work. Research at the Perimeter Institute is supported by the Government of Canada through Industry Canada and by the Province of Ontario through the Ministry of Research \& Innovation. LB also acknowledges support from an NSERC Discovery Grant. SF acknowledges support from the Max Planck Society.  

\appendix

\section{More about $\ast$-DGAs}
\label{dga_appendix}

This appendix is meant to help the unfamiliar reader become better acquainted with some of the basic rules defining $\ast$-DGAs that we introduced in Section~\ref{dga} -- the logic of how they interact with one another, why they are what they are, and how much freedom there is to modify them.

\subsection{$*$-DGA conventions and definitions}
\label{graded_Leibniz_rule}

In Section~\ref{dga} we built up the defining properties of $*$-DGAs in steps, first introducing the involution $*$, followed by the grading $A = \bigoplus_n A^n$, and finally introducing the differential operators $d_L$ and $d_R$. In this section we wish to highlight how some of these conditions arise, and where matters of convention enter. To do so we consider generalized $*$-DGAs which satisfy the following generalizations of the conditions given in \eqref{graded_Leibniz} and Eqs.~\eqref{graded_star}: 
\begin{subequations}
\label{DGA_Compat_ALL}
\begin{eqnarray}
  \label{DGA_Compat_Leib}
  d_{L}(a_{m}a_{n})&=&\rho_n d_{L}(a_{m})a_{n}+\eta_{m}a_{m}d_{L}(a_{n}),\\
  \label{DGA_Compat_chi}
  (a_ma_n)^* &=&\chi_{m,n}a_n^*a_m^*, 
\end{eqnarray} 
\end{subequations}
for $a_m,a_n\in A$, and where the coefficients $\rho_n,\eta_{m},$ and $\chi_{m,n}$ are valued in $\mathbb{F}$. We maintain the usual conditions $d^2=0$ and $(a^*)^*= a$. Our goal will be  to understand what restrictions are forced on the coefficients in Eqs.~\eqref{DGA_Compat_ALL}. 

\subsubsection{The graded Leibniz rule}

Let's start by considering the coefficents $\rho_n,\eta_m$ in Eq.~\eqref{DGA_Compat_Leib}. By applying $d_L$ twice to a pair of algebra elements $a_ma_n$, and making use of Eq.~\eqref{DGA_Compat_Leib} we find
\begin{equation}
  \label{dL2}
  d_{L}^{2}(a_{m}a_{n})=d_{L}^{2}(a_{m})a_{n}+(\rho_n\eta_{m+\epsilon''}+\rho_{n+\epsilon''}\eta_{m})d_{L}(a_{m})d_{L}(a_{n})
  +\eta_{m}^{2}a_{m}d_{L}^{2}(a_{n}).
\end{equation}
Our first requirement for a DGA is that we want $d_{L}^{2}=0$ on $A$, and from (\ref{dL2}) we see that this implies $\eta_{m}=(-\rho_{n+1}/\rho_n)^m\eta_{0}$, and $\rho_{m}=(-\eta_{n+1}/\eta_n)^m\rho_{0}$. This  implies that $\rho$, and $\eta$ are of the form: $\rho_m = \kappa^m \rho_0,$ and $\eta_m =  (-\kappa)^m\eta_0$ for some $\kappa,\eta_{0},\rho_0\in\mathbb{F}$.  To further fix $\eta_{0}=\rho_0=1$, we can use one of the following three requirements -- any one of them will do the job.  First, we could require that 
$d_{L}$ should obey the {\it ordinary} (ungraded) Leibniz rule on the sub-algebra of zero forms $A_{0}$:
\begin{equation}
  d_{L}(a_{0}a_{0}')=d_{L}(a_{0})a_{0}'+a_{0}d_{L}(a_{0}'),
\end{equation}
which directly fixes $\eta_{0}=\rho_0=1$.  Second, we could require that the Leibniz rule is the one that makes sense for a unital algebra $A$ (with unit $e\in A_{0}$), in which case
\begin{equation}
  d(a_{0})=d(e a_{0})=d(e)a_{0} + \eta_{0} e d(a_{0}) = \eta_{0} d(a_{0})
\end{equation}
which again fixes $\eta_{0}=1$. An analogous argument from the right fixes $\rho_0=1$.  Third, we could require that the Leibniz rule is the one that makes sense for an associative algebra $A$, so that it does not matter in which order we use the Leibniz rule to break up the expression $d(a_{m}a_{n}a_{p})$ into parts; this implies $\eta_{m}\eta_{n}=\eta_{m+n}$ which, once again, implies $\eta_{0}=\rho_0=1$.  All roads lead in the same direction, and so we have:
\begin{align}
\rho_m = \kappa^m ,\hspace{2cm}\eta_m =  (-\kappa)^m\label{dga_leib_kappas}
\end{align}

Next notice that given a differential operator $d_L$ satisfying the generalized Leibniz rule given in eq~\eqref{DGA_Compat_Leib} and~\eqref{dga_leib_kappas}, we can always define a new differential operator $d_L'(a_m) = (\pm \kappa)^{-m}d_L(a_m)$ which will satisfy the graded Leibniz rule $d_L'(a_m a_n) = (\pm1)^n d_L'(a_m)a_n + (\mp)^m a_m d_L'(a_n)$. Without any loss of generality we therefore choose the simplest case $\kappa = 1$, in eq~\eqref{dga_leib_kappas} such that we arrive at the `left' graded Leibniz rule (\ref{graded_Leibniz}). It is called the `left' graded Leibniz rule because signs are picked up whenever passing over an algebra element from the left. An equally good choice would have been to select $\kappa = -1$, which would have resulted in a `right' acting differential operator. 

\subsubsection{Properties of graded involutions}
\label{DGA_invol_prop}
Let us next consider what role the function $\chi_{m,n}$ plays in our construction.  In Eq.~\eqref{dga_dR} we introduced $d_R = * d_L\overline{*}$ as a `right-acting' differential. Notice however that this interpretation depended on the form of $\chi_{m,n}$. In particular, for a $*$-DGA satisfying  the generalized condition~\eqref{DGA_Compat_ALL}, $d_R$ satisfies the generalized Leibniz rule:
\begin{align}
d_R(a_m a_n) =(-1)^n\overline{\chi_{m,n}}\chi_{\epsilon''n,\epsilon''m+1}d_R(a_m)a_n+ \overline{\chi_{m,n}}\chi_{\epsilon''n+1,\epsilon''m}a_m d_R(a_n).
\end{align}
We see that our choice $\chi_{m,n} = 1$ in Section~\ref{dga} indeed results in $d_R = *d_L \overline{*}$ acting as a right differential; but what freedom did we have in this choice? Our first constraint is that we require $(a_m^*)^* = a_m$, which implies $\chi_{m,n} = (\overline{\chi_{\epsilon''n,\epsilon''m}})^{-1}$. Then, for the involution to make sense for associative algebras we need $\chi_{m,n+p}\chi_{n,p} = \chi_{m,n}\chi_{m+n,p}$. A choice compatible with both of  these restrictions is to select $\chi_{m,n}=(-1)^{mn}$; but for that choice the operator $d_R = *\circ d_L\circ \overline{*}$ acts as a `left' differential, and it  is  instead the operator $d'_R(a_m) = (-1)^m *d_L\overline{*}(a_m)$ which acts as a right differential. Notice that these two conventional choices ($\chi_{m,n} = 1$ or $\chi_{m,n} = (-1)^{mn}$) work for both real and complex $*$-DGAs, and for $*$-DGAs of both types ($\epsilon'' = \pm 1$). Both conventions turn out to be very useful in different situations as we will show below in Subsection~\ref{graded_product_rule}. 
More generally, we can write
\begin{align}
d_R = \sigma_m' *d_L\overline{*},\label{dga_compat}
\end{align}
where $\chi_{mn} = e^{i\alpha mn}$, $\sigma_m' = \pm e^{-i\alpha m}$, and we will refer to the special cases $\alpha=0$ and $\alpha=1$ as "convention I" and "convention II", respectively.\footnote{As discussed in Subsection~\ref{defining_A}, for $*$-DGAs satisfying $\epsilon'' = 1$, we could impose the additional condition $d_L(a_m^*) = \pm e^{i(\pi - \alpha)m} d_L(a_m)^*$, or equivalently using Eq.~\eqref{dga_compat}, $d_L(a_m) = \pm(-1)^m d_R(a_m)$. In this paper we do not enforce this additional condition however.} For complex DGAs satisfying $\epsilon'' = 1$, we could consider the more general situation in which $-\pi\le\alpha\le \pi$. Notice however that whichever value of $\alpha$ is chosen, we can always construct a new involution $*'$ which is given by $a_m^{*'} = e^{i\tau (m-1)m/2}(a_m)^*$. For this new involution $*'$ one finds a new coefficient $\chi_{m,n}' = \chi_{m,n}e^{i\tau mn}$.  Thus we can always pick $\tau$ to recover either convention I or II.
 
\subsection{Graded tensor product conventions and definitions}
\label{graded_product_rule}

Next let us think about the origin of the (Kozul) tensor product rule given in (\ref{tensor_product}). Suppose $H'$ and $H''$ are graded vector spaces on which graded operators $\mathcal{O}_m':H_p'\rightarrow H_{p+m}'$ and $\mathcal{O}_n'':H_q''\rightarrow H_{q+n}''$ act respectively (where the subscripts denote the order of each operator). Define the product space  $H=H'\otimes H''$ and product operator $\mathcal{O}_{m+n} =\mathcal{O}_{m}'\otimes \mathcal{O}_{n}''$ such that the action of $\mathcal{O}_{m+n}$ on elements of $H$ is given by:
\begin{equation}
  (\mathcal{O}_{m}'\otimes \mathcal{O}_{n}'')(h_{p}'\otimes h_{q}'')=\Psi_{n,p}(\mathcal{O}_{m}'h_{p}'\otimes \mathcal{O}_{n}''h_{q}'')\qquad(\chi_{n,p}\in\mathbb{F})\label{dga_Kozul}
\end{equation}
for $h_p'\in H_p'$, $h_q''\in H_q''$, and where $\Psi_{n,p}$ are $\mathbb{F}$ valued coefficients that we wish to constrain.  Note that $\ast$-DGAs can be thought of as particular examples of graded vector spaces in which each element  $a_m\in A_m$ can be considered as an operator of degree $m$. Given two $*$-DGAs $(A',d',*')$, $(A'',d'',*'')$, define their tensor $A=A'\otimes A''$ following Eq.~\eqref{dga_Kozul} such that the product between algebra elements is given by
\begin{subequations}
\begin{equation}
  (a_{m}'\otimes a_{n}'')(a_{p}'\otimes a_{q}'')=\Psi_{n,p}(a_{m}'a_{p}'\otimes a_{n}''a_{q}'')\qquad(\chi_{n,p}\in\mathbb{F})
\end{equation}
and where the differential and involution are given by 
\begin{align}
  d&=d\;\!{}'\otimes 1''+1'\otimes d\;\!{}'',\label{graded_diff}\\ 
* &= *'\otimes *'' \theta,
\end{align}
\end{subequations}
where $\theta$ is an $\mathbb{F}$ valued function of the grading, which again we wish to determine.

Let's first determine the form of the function $\Psi_{n,m}$. We want $(A,d)$ to be a DGA ({\it i.e.}\ we want $d$ to satisfy the appropriate left graded Leibniz rule on $A$), which implies $\Psi_{n,p}=(-1)^{np}\Psi_{0,0}$\footnote{Note that we could have instead asked that an appropriate right-leibniz rule be satisfied in which case we would have arrived at a different convention for the Kozul sign.}.  This, in turn, is enough to ensure that:
\begin{itemize}
\item (i) {\it if} $A'$ and $A''$ are both associative, then $A$ is also associative; and
\item (ii) {\it if} $A'$ and $A''$ are both graded-commutative, then $A$ is also graded-commutative.
\end{itemize}
(We stress that we do {\it not} assume associativity or graded-commutativity in this appendix.)

In order to further fix $\Psi_{0,0}=1$, we can use one of the following two requirements -- either one will do the job.  First, we could require that, on the sub-algebra of zero-forms $A_{0}'\otimes A_{0}''$, the multiplication rule should reduce to the standard one for the tensor product of two ungraded algebras $A_{0}'$ and $A_{0}''$:
\begin{equation}
  (a_{0}'\otimes a_{0}'')(\tilde{a}_{0}'\otimes\tilde{a}_{0}'')=a_{0}'\tilde{a}_{0}'\otimes a_{0}''\tilde{a}_{0}''
\end{equation} 
which directly fixes $\Psi_{0,0}=1$.  Second, we could require that the product is the one that makes sense for unital algebras $A'$, $A''$ and $A'\otimes A''$ (with units $e'$, $e''$ and $e'\otimes e''$, respectively), so that
\begin{equation}
  (a_{0}'\otimes a_{0}'')=(e'\otimes e'')(a_{0}'\otimes a_{0}'')=\chi_{0,0}(e'a_{0}'\otimes e''a_{0}'')=\chi_{0,0}(a_{0}'\otimes a_{0}'')
\end{equation}
which also fixes $\Psi_{0,0}=+1$.  Either path leads back to the (Kozul) product rule (\ref{tensor_product}).

Next let us consider the function $\theta$. In Subsection~\ref{DGA_invol_prop} we introduced two good conventions for defining the involution on a differential graded algebra labelled by the signs $\chi_{m,n} = e^{i\alpha mn}$, $\sigma_m' = \pm e^{-i\alpha m}$ for $\alpha = 0$ or $\alpha = \pi$. Once a convention is chosen, if we want it to remain stable under the tensor product of $*$-DGAs, then this places restrictions on the function $\theta$. In particular,
it should be:
\begin{align}
\theta(a_m\otimes a_n)=e^{i(\pi-\alpha)mn}(a_m\otimes a_n)
\end{align}
with $\alpha = 0$ (for convention I) or $\alpha = \pi$ (for convention II). The  convention chosen in the body of this paper corresponds to the choice $\chi_{m,n} = \sigma'_m = 1,$ and $\theta_{m,n}=(-1)^{mn}$. 

In future work, we will discuss the implications of this formalism for taking tensor products of spectral triples, since this is an interesting story in its own right \cite{Farnsworth:2016qbp}.


\end{document}